\documentclass[fleqn,usenatbib,useAMS]{mnras}

\usepackage{graphicx}	
\usepackage{amsmath}	
\usepackage{amssymb}	
\usepackage{multicol}        
\usepackage{bm}		
\usepackage{pdflscape}	
\usepackage{array}

\usepackage[T1]{fontenc}
\usepackage{ae,aecompl}

\usepackage{newtxtext,newtxmath}

\DeclareRobustCommand{\VAN}[3]{#2}
\let\VANthebibliography\thebibliography
\def\thebibliography{\DeclareRobustCommand{\VAN}[3]{##3}\VANthebibliography}

\newcommand{\Msun}{\rm M_{\odot}}

\newcommand{\HII}{{H}{\sc {ii}}}
\newcommand{\mm}{$\mu$m}

\title[Radio spectral properties of SFGs between 150--5,000\,MHz]{Radio spectral properties of star-forming galaxies between 150--5,000\,MHz in the ELAIS-N1 field}

\author[Fangxia An et al.]{
Fangxia An,$^{1,2}$\thanks{E-mail: fangxiaan@pmo.ac.cn, fangxiaan@gmail.com}
M. Vaccari,$^{3,2,4}$
P. N. Best,$^{5}$
E. F. Ocran,$^{6}$
C. H. Ishwara-Chandra,$^{7,3}$
A. R. Taylor,$^{3,2}$
\newauthor
S. K. Leslie,$^{8}$
H. J. A. R{\"o}ttgering,$^{8}$
R. Kondapally,$^{5}$
Paul Haskell,$^{9}$
J. D. Collier,$^{3,10}$
M. Bonato$^{4,11}$
\\
$^{1}$Purple Mountain Observatory, Chinese Academy of Sciences, 10 Yuanhua Road, Qixia District, Nanjing 210023, People's Republic of China\\
$^{2}$Inter-University Institute for Data Intensive Astronomy, and Department of Physics and Astronomy, University of the Western Cape, Robert Sobukwe Road,\\ 7535 Bellville, Cape Town, South Africa\\
$^{3}$Inter-University Institute for Data Intensive Astronomy, and Department of Astronomy, University of Cape Town, Private Bag X3, Rondebosch 7701, South Africa\\
$^{4}$INAF-Istituto di Radioastronomia, via Gobetti 101, 40129 Bologna, Italy\\
$^{5}$Institute for Astronomy, University of Edinburgh, Royal Observatory, Blackford Hill, Edinburgh, EH9 3HJ, UK\\
$^{6}$Korea Astronomy and Space Science Institute, 776 Daedeokdae-ro, Daejeon 305-348, Korea\\
$^{7}$National Centre for Radio Astrophysics, Tata Institute of Fundamental Research, Pune 411007, India\\
$^{8}$Leiden Observatory, Leiden University, PO Box 9513, NL- 2300 RA Leiden, The Netherlands\\
$^{9}$Centre for Astrophysics Research, University of Hertfordshire, Hatfield, AL10 9AB, UK\\
$^{10}$School of Science, Western Sydney University, Locked Bag 1797, Penrith, NSW 2751, Australia\\
$^{11}$Italian ALMA Regional Centre, Via Gobetti 101, 40129 Bologna, Italy\\
}
\date{Accepted XXX. Received YYY; in original form ZZZ}

\pubyear{2024}

\begin{document}
\label{firstpage}
\pagerange{\pageref{firstpage}--\pageref{lastpage}}
\maketitle

\begin{abstract}
By combining high-sensitivity LOFAR 150\,MHz, uGMRT 400\,MHz and 1,250\,MHz, GMRT 610\,MHz, and VLA 5\,GHz data in the ELAIS-N1 field, we study the radio spectral properties of radio-detected star-forming galaxies (SFGs) at observer-frame frequencies of 150--5,000\,MHz. We select $\sim$3,500 SFGs that have both LOFAR 150\,MHz and GMRT 610\,MHz detections, and obtain a median two-point spectral index of $\alpha_{150}^{610}=-0.51\pm0.01$. The photometric redshift of these SFGs spans $z=0.01-6.21$. We also measure the two-point radio spectral indices at 150--400--610--1,250\,MHz and 150--610--5,000\,MHz respectively for the GMRT 610-MHz-detected SFGs, and find that, on average, the radio spectrum of SFGs is flatter at low frequency than at high frequency. At observer-frame 150--5,000\,MHz, we find that the radio spectrum slightly steepens with increasing stellar mass. However, we only find that the radio spectrum flattens with increasing optical depth at $V$-band at $\nu\la1$\,GHz. We suggest that spectral ageing due to the energy loss of CR electrons and thermal free-free absorption could be among the possible main physical mechanisms that drive the above two correlations respectively. In addition, both of these mechanisms could physically explain why the radio spectrum is flatter at low frequency than at high frequency.
\end{abstract}




\section{Introduction} \label{sec:intro}
The radio continuum emission from star-forming galaxies (SFGs) consists of non-thermal synchrotron radiation and thermal free-free emission. In SFGs, massive stars ($M>8\,\Msun$) explode to form Type II and Type Ib supernovae and accelerate cosmic ray (CR) electrons, which gyrate within galactic magnetic fields and produce radio emission via a synchrotron process \citep{Voelk89}. The underlying physical process of free-free emission is powered by the Coulomb scattering between free ions and electrons in the \HII\ regions, which are ionised by young, massive stars. Therefore, both of these processes are associated with massive star formation in galaxies. These characteristics, as well as the radio emission's insensitivity to dust attenuation, make the radio continuum a powerful tool to study the astrophysical properties of SFGs \citep[see][for a review]{Condon92}. 

Although the free-free emission traces a nearly instantaneous star formation \citep[$\la10$\,Myr,][]{Murphy11, Murphy12, Kennicutt12}, it is challenging to target at gigahertz frequencies or below due to its flat spectrum ($\alpha\sim-0.1$, where $S_{\nu}\propto\nu^{\alpha}$). At $\nu\ga$30\,GHz, thermal free-free radiation is expected to dominate the radio continuum emission, but the observation condition at such high-frequency range is stringent \citep{Murphy11, Murphy12, Algera21, Algera22}. Therefore, although it traces integrated star formation in the last $\sim$100\,Myr, the non-thermal synchrotron emission, which has a steep spectrum of $\alpha\sim-0.8$, is widely used to study activities of SFGs in both the local and distance Universe \citep[e.g.,][]{Condon92, Bressan02,Galvin16,Matthews21,Perez21,Arango23,Vollmer23}. 

On galaxy-integrated scales, the radio-continuum-based star formation rate (SFR) is most often calibrated by an empirical relation between far-infrared (FIR) and radio (typically at $\sim$1.4\,GHz) luminosities known as the FIR radio correlation (FIRRC). The FIRRC appears to hold across several orders of magnitude in spatial scale, luminosity, gas surface density, and photon/magnetic field density, while it demonstrates a modest evolution with redshift and potentially an evolution with stellar mass \citep[e.g.,][]{Helou85, Yun01, Bell03, Ivison10, Mao11, Magnelli15, Hindson18, Delvecchio21,Smith21}. In addition, the FIRRC is also widely used to classify radio sources as either SFGs or radio-loud active galactic nuclei (AGN), because it is characteristic of radio-loud AGN activities to accelerate the CR electrons and cause an `excess' of radio emission \citep[e.g.,][]{Smolcic17,Delvecchio21, Whittam22}. Therefore, studies of SFGs based on radio continuum emission rely on a well-determined radio spectrum, especially at low frequency since the rest-frame $\sim1.4$\,GHz is shifted to $\nu<1.4$\,GHz at high redshift. 

However, due to the lack (until recently) of high-sensitivity and large coverage low-frequency radio surveys and the fact that the radio spectra of SFGs are dominated by synchrotron radiation at $\nu\sim1-10$\,GHz, most of the studies based on rest-frame radio spectra of SFGs assumed a synchrotron dominated spectrum with a spectral index of $\alpha\sim(-0.7)$--($-0.8)$ when $k$-correcting the observed flux densities at $\sim1.4$\,GHz \citep[e.g.,][]{Magnelli15, Hindson18, Delvecchio21}. These assumptions will overestimate the radio flux density at low frequency because of the flattening of radio spectrum at $\nu\la1$\,GHz as suggested by some recent works \citep{Schleicher13, Delhaize17, Galvin18, Gim19, An21,Bonato21}. 

%
%
\begin{table*}
\caption{Summary of the radio data sets used in this work}
\label{tab:table1}
\begin{tabular}{lccccccc}
\hline
 Data sets  & Effective frequency & Bandwidth & Coverage & Minimum RMS noise & Median RMS noise &  Resolution & Number of sources \\
 & MHz & MHz & deg$^{2}$ & $\mu$Jy\,beam$^{-1}$ & $\mu$Jy\,beam$^{-1}$ & arcsecond ($''$)  \\
\hline
 LOFAR 150\,MHz  & 146.2  & 62& 6.7  & 17.1 & 22.9  & 6 & 31,610 \\
 uGMRT 400\,MHz  & 400 & 200 & 1.8  & $\sim$15 &  & $4.6\times4.3$ & 2,528 \\
 GMRT-deep 610\,MHz  & 612 & 32  & 1.9  & 7.1 & 19.5  & 6 & 4,290 \\
 GMRT-wide 610\,MHz  & 612 & 32  & 12.8 &  & $\sim$40  & 6 & 6,400 \\
 uGMRT 1,250\,MHz & 1250 & 400 & 2.5$^a$ & $\sim$12 & & 2.3$\times$1.9 & 1,086 \\
 VLA 5\,GHz  & 5,000 & 2,000  & 0.1  &  & 1.1 & 2.5 & 387 \\
 \hline
 \end{tabular}
 \\
 $^a$ Source detection in \cite{Sinha23} was carried out within the central 1.1\,deg$^{2}$ .\\
\end{table*}

At frequency $\nu\la1$\,GHz, besides the synchrotron and thermal free-free emission processes, the spectra of SFGs are expected to be affected by additional physical mechanisms. Notably, the \HII\ regions become optically thick at $\nu\la1$\,GHz, where free-free absorption causes greater attenuation of lower-frequency synchrotron radiation. This effect has been observed in the Galactic centre and some nearby galaxies \citep{Wills97, Roy13, Varenius15}, and it has been widely used to explain a flattening of the radio spectrum at low frequency for both local and high-redshift galaxies \citep[e.g.,][]{Condon92, McDonald02, Murphy09, Clemens10, Lacki13, Galvin18, Klein18, Dey22}. However, some other studies suggest that thermal gas absorption might not be the main cause of a flattening of the radio spectrum at low frequency, since, for instance, there is no correlation between the inclination of the galaxies and the flatness of their radio spectrum at low frequency \citep{Marvil15, Chyzy18}. Alternatively, the intrinsic steepening of the synchrotron spectrum, due to synchrotron and inverse Compton losses growing stronger at high CR energy, {has been} used to explain why the high-frequency radio spectrum is steeper than the low-frequency one \citep{Marvil15, Chyzy18,Thomson19, An21,Sweijen22}. Besides, the Razin-Tsytorich effect \citep{Razin60}, bremsstrahlung and ionisation losses of CR electrons, synchrotron self-absorption, and free-free emission could flatten the radio continuum spectrum at low frequency \citep[e.g.,][]{Fleishman95,Murphy09,Lacki10,Lacki13,Basu15,Heesen22}.

In recent years, the high-sensitivity and wide-field extragalactic radio continuum surveys conducted by the Low Frequency Array \citep[LOFAR,][]{van13}, Giant Metrewave Radio Telescope \citep[GMRT,][]{Swarup91}, and Murchison Widefield Array \citep[(MWA,][]{Lonsdale09} are finally allowing us to study the radio spectral properties at $\nu_{\rm obs}\la1$\,GHz with a statistically significant sample of SFGs. These studies become increasingly important as radio continuum observations are expected to become more powerful in exploring cosmic evolution with forthcoming facilities such as the Square Kilometre Array \citep[SKA,][]{Dewdney09} and next-generation $Karl\,G.\,Jansky$ Very Large Array \citep[ngVLA,][]{Murphy18}.

In this work, we combine high-sensitivity LOFAR 150\,MHz data, GMRT 610\,MHz data, 400\,MHz and 1,250\,MHz data from the upgraded GMRT (uGMRT), and 5\,GHz data observed by the VLA in the ELAIS-N1 field \citep[European Large Area ISO Survey North 1,][]{Oliver00} to study radio spectral properties at observer-frame frequencies of 150--5,000\,MHz. Details of these radio data as well as the ancillary data used in this work are described in Section $\S$\ref{s:observation}. We also summarise the main properties of these radio datasets in Table \ref{tab:table1}. We analyse the data and select SFGs from the radio surveys in Section $\S$\ref{s:Analysis}. The measured radio spectral indices at different frequency ranges and the correlations between radio spectrum and physical properties of SFGs are presented in Section $\S$\ref{s:results}. Using these results, we discuss the underlying physical mechanisms that drive the radio spectrum at rest-frame $\sim$150--10,000\,MHz in Section $\S$\ref{s:discussion}. Our results are summarised in Section $\S$\ref{s:conclusion}.

Throughout this paper, we adopt the AB magnitude system \citep{Oke74} and assume a flat $\Lambda$CDM cosmological model with the Hubble constant $H_0 = 67.27$\,km\,s$^{-1}$\,Mpc$^{-1}$, matter density parameter $\Omega_{\rm m} = 0.32$, and cosmological constant $\Omega_{\Lambda} = 0.68$ \citep{Planck16}.
\section{Observations and data} \label{s:observation}

\subsection{LOFAR 150\,MHz data} \label{s:lofar_obs}

The LOFAR 150\,MHz data of the ELAIS-N1 field used in this work are from the  LOFAR Two Metre Sky Survey \citep[LoTSS,][]{Shimwell17} Deep Fields \citep{Tasse21, Sabater21}. The observations were taken with the High Band Antenna (HBA) array with frequencies between 115--177\,MHz in the LOFAR observation cycles 0, 2, and 4. The central frequency of the observation is 146.2\,MHz. The final Stokes $I$ image for all survey fields has a resolution of 6$\arcsec$. The ELAIS-N1 field is the deepest of the LoTSS Deep Fields to date with an effective observing time of 163.7\,hours \citep{Sabater21}. The median RMS of the central 6.7\,deg$^{2}$, where the highest-quality multi-wavelength data exists \citep[from 0.15 to 500\,\mm,][]{Kondapally21}, is 22.9\,$\mu$Jy\,beam$^{-1}$. Within this region, \cite{Kondapally21} created a final cross-matched and associated catalogue that contains 31,610 LOFAR 150\,MHz radio sources (Figure~\ref{f:coverage}). We refer the reader to \cite{Sabater21} for more details about the observations and data reduction of LOFAR 150\,MHz data and to \cite{Kondapally21} for the creation of the final cross-matched and associated radio catalogue. We also present a short summary of the cross-matched results in Section $\S$\ref{s:SFGs_selection_150}.

\subsection{uGMRT 400\,MHz data} \label{s:ugmrt}
\begin{figure}
\centering
\includegraphics[width=0.48\textwidth]{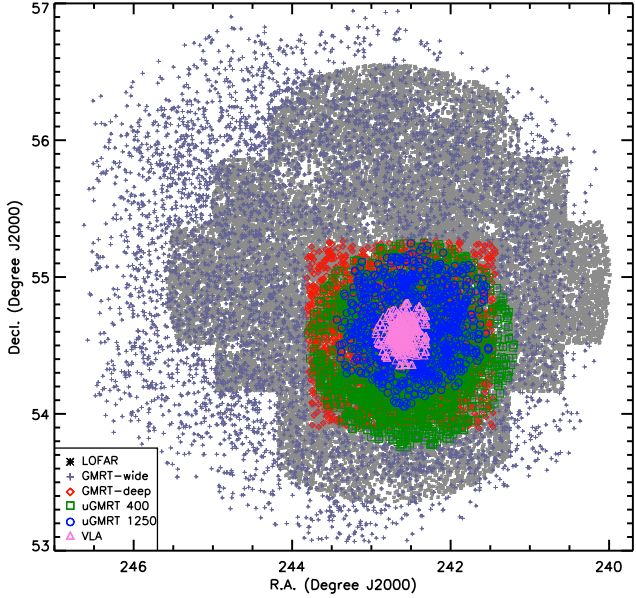}
\caption{Distribution of radio sources detected at LOFAR 150\,MHz, uGMRT 400 and 1,250\,MHz, GMRT 610\,MHz, and VLA 5\,GHz respectively in the ELAIS-N1 field.} 
\label{f:coverage}
\end{figure}

The ELAIS-N1 field was observed at 300--500\,MHz with the uGMRT in GTAC cycle 32 \citep{Chakraborty19}. The observations cover an area of $\sim$1.8\,deg$^{2}$. With a total on-source time of $\sim$13\,hours, the observations reach an off-source RMS noise of $\sim$15\,$\mu$Jy\,beam$^{-1}$ near the centre of the field, which is the minimum RMS noise of the uGMRT 400\,MHz data as listed in Table~\ref{tab:table1}. The angular resolution of the image is $4\farcs6\times4\farcs3$. \cite{Chakraborty19} detected 2,528 radio sources at $>6\,\sigma$ from the primary beam-corrected uGMRT 400\,MHz image (Figure~\ref{f:coverage}). Details of observation, data reduction, and source detection are presented in \cite{Chakraborty19}. 

%
%

\subsection{GMRT 610\,MHz data} \label{s:gmrt}
There are two sets of GMRT 610\,MHz data of the ELAIS-N1 field used in this work. One is from the ELAIS-N1 610\,MHz Deep Survey \citep[GMRT-deep,][]{Ocran20}. The other is from a wide-area GMRT 610\,MHz survey of the ELAIS-N1 field \citep[GMRT-wide,][]{Ishwara20}. The GMRT-deep data include a set of pointings from the GMRT-wide survey \citep{Ocran20}.

\subsubsection{GMRT-deep}\label{s:gmrt_deep}
The ELAIS-N1 610\,MHz Deep Survey covers an area of 1.86\,deg$^{2}$ in the centre of the field. The survey was carried out with seven closely spaced GMRT pointings with an on-source integration time of $\sim$18\,hours for each pointing \citep{Ocran20}. The seven pointings were mosaicked to create the final image. To reduce the noise around the edge, \cite{Ocran20} also includes a set of pointings with on-source time of 2.5\,hours per pointing from GMRT-wide survey in the final mosaic image. The median RMS noise of the final mosaic image is 19.5\,$\mu$Jy\,beam$^{-1}$ with a minimum noise of 7.1\,$\mu$Jy\,beam$^{-1}$ in the central region. The angular resolution of the GMRT-deep image is 6$\arcsec$. 
\cite{Ocran20} detected 4,290 $>5\,\sigma$ radio sources from the final mosaic image (Figure~\ref{f:coverage}). We refer the reader to \cite{Ocran20} for more details. 
 
\subsubsection{GMRT-wide}\label{s:gmrt_wide}
The wide-area GMRT 610\,MHz survey covers an area of 12.8\,deg$^{2}$ in the ELAIS-N1 field by using a total of 51 GMRT pointings. The on-source time for each pointing is $\sim$2.5\,hours and for the final mosaic data is 153\,hours. The final mosaic image reaches an RMS noise of $\sim$40\,$\mu$Jy\,beam$^{-1}$ and an angular resolution of 6$\arcsec$. A total of 6,400 radio sources were detected from the final mosaic image (Figure~\ref{f:coverage}). Details of observations, data reduction, and source extraction are presented in \cite{Ishwara20}.

\subsection{uGMRT 1,250\,MHz data} \label{s:ugmrt_1250}
We use the uGMRT 1,250\,MHz (L-band) data from \cite{Sinha23} in this work. The bandwidth of the uGMRT L-band is 400\,MHz. The final mosaic image covers the central $\sim$2.4\,deg$^{2}$ of the ELAIS-N1 field and reaches a minimum central off-source RMS noise of 12\,$\mu$Jy\,beam$^{-1}$ with a beam size of $2\farcs3\times1\farcs9$. A total of 1,086 were detected above $5\,\sigma$ within the central $\sim$1.1\,deg$^{2}$ with a flux limitation of $S_{\rm 1250}>60\,\mu$Jy \citep{Sinha23}. We refer the reader to \cite{Sinha23} for details of the observation, data reduction, and source detection.

\subsection{VLA 5\,GHz data} \label{s:JVLA}
The VLA 5\,GHz observations only cover the central 0.13\,deg$^{2}$ of the ELAIS-N1 field \citep{Taylor14}. The bandwidth of the observation is 2\,GHz and the central frequency is 5\,GHz. With a total of 60 hours observational time, the VLA 5\,GHz data reaches an RMS noise of 1.05\,$\mu$Jy\,beam$^{-1}$. The angular resolution of the image is $2\farcs5$. In total, \cite{Taylor14} detected 387 radio sources with $S_{\rm 5GHz}>5\,\mu$Jy from the image. Although \cite{Taylor14} only focused on the polarized emission from these sources, we include their continuum emission in this work to measure the radio spectral index between the observer-frame frequencies of 150--5,000\,MHz.

\subsection{Additional Multi-wavelength Data/Catalogues}\label{s:other data}
The ELAIS-N1 field was originally chosen for deep extra-galactic observations with the Infrared Space Observatory (ISO) due to its low infrared background \citep{Rowan04, Vaccari05}. The deep infrared, as well as the ultraviolet (UV) and optical observations, make this field one of the best-studied astronomical fields. Since this work is based on the radio-selected SFGs, we do not use these multi-wavelength data directly but utilize the cross-matched results of GMRT 610\,MHz radio sources from \cite{Ocran20} and of LOFAR 150\,MHz sources from \cite{Kondapally21}. We therefore give a short summary of the multi-wavelength data used in the cross-matches in their works and refer the reader to \cite{Ocran20} and \cite{Kondapally21} for more details of these data. 

\cite{Ocran20} determined the multi-wavelength counterparts of GMRT 610\,MHz-detected radio sources by using the Spitzer Extragalactic Representative Volume Survey (SERVS) DR2 firstly \citep{Mauduit12, Vaccari15}. The SERVS Data Fusion project included most of the publicly available far-UV to FIR data and homogenized these data to estimate the photometric and spectroscopic redshifts for IRAC 3.6\,\mm\ and/or 4.5\,\mm-selected sources \citep{Vaccari15}. 
Furthermore, if a SERVS match was not found, \cite{Ocran20} used the $J$ and $K$-band data from the UKIRT Infrared Deep Sky Survey (UKIDSS) Deep Extragalactic Survey (DXS) \citep[DR10,][]{Lawrence07} to identify counterparts to GMRT 610\,MHz radio sources. We will be back to their cross-matched results in Section $\S$\ref{s:Analysis}.

In addition to the deep 3.6\,\mm\ and 4.5\,\mm\ from the SERVS and near-infrared (NIR) $J$ and $K$-band data from UKIDSS-DIS, the following datasets were also used by \cite{Kondapally21}:
\begin{enumerate}
\item UV data from the Deep Imaging Survey (DIS) taken with the Galaxy Evolution Explorer (GALEX) space telescope \citep{Martin05, Morrissey07},
\item $u$-band data from the Spitzer Adaptation of the Red-sequence Cluster Survey \citep[SpARCS,][]{Wilson09, Muzzin09},
\item optical $g$, $r$, $i$, $z$, and $y$-band data from the Medium Deep Survey (MDS), one of the Panoramic Survey Telescope and Rapid Response System (Pan-STARRS-1) surveys \citep{Chambers16},
\item the $g$, $r$, $i$, $z$, $y$-band, and the narrowband NB921 data from the Hyper-Suprime-Cam Subaru Strategic Program (HSC-SSP) survey \citep{Aihara18}, and
\item the mid-infrared (MIR) 3.6\,\mm, 4.5\,\mm, 5.8\,\mm, and 8.0\,\mm\ data from the Spitzer Wide-area Infra-Red Extragalactic (SWIRE) survey \citep{Lonsdale03}.
\end{enumerate}
In their multi-wavelength photometric catalogue that was used to cross-match with the LOFAR 150\,MHz radio sources, \cite{Kondapally21} included the far-infrared (FIR) data at 24\,\mm\ from the Multi-band Imaging Photometer for Spitzer \citep[][]{Rieke04} and at 100\,\mm, 160\,\mm, 250\,\mm, 360\,\mm, and 520\,\mm\ from Herschel Multi-tiered Extragalactic survey \citep[HerMES,][]{Oliver12}. We will introduce their cross-matched results in Section $\S$\ref{s:Analysis}.

\section{Analysis}\label{s:Analysis}
To investigate the radio spectral properties of radio-detected sources, we first check the consistency of the astrometry and flux density calibration of the radio datasets used in this work. Then, we introduce the selection of SFGs from GMRT 610\,MHz-detected sources in \cite{Ocran20} and from LOFAR 150\,MHz-detected sources in \cite{Best23}. 

%
%
\begin{figure*}
\centering
\includegraphics[width=0.98\textwidth]{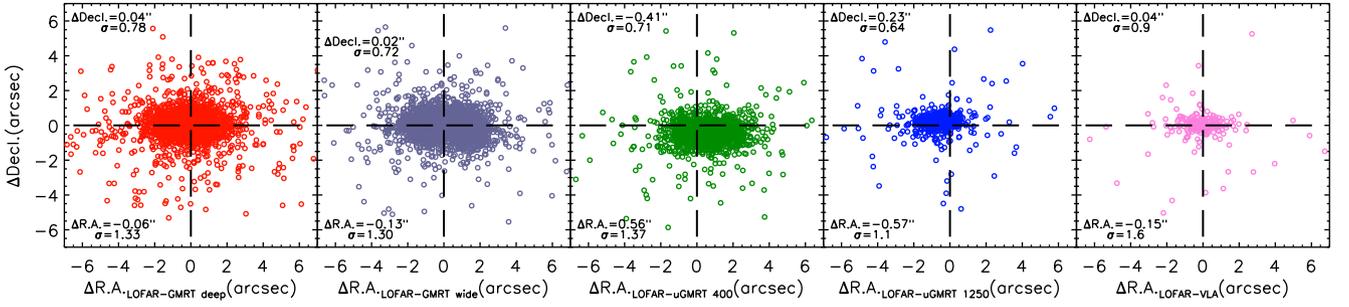}
\caption{Comparisons between positions of sources detected in LOFAR 150\,MHz, uGMRT 400\,MHz and 1,250\,MHz, GMRT 610\,MHz (both GMRT-deep and GMRT-wide), and VLA 5\,GHz images. Each pair of catalogues is matched by using a matching radius of 10\,$\arcsec$. The astrometry of LOFAR 150\,MHz sources is well-consistent with that of GMRT 610\,MHz sources (from both deep and wide images) and VLA 5\,GHz as shown in the first two panels and the last panel respectively. The astrometric offsets between LOFAR 150\,MHz-detected sources and uGMRT 400\,MHz, and 1,250\,MHz sources are relatively larger as shown in the third and fourth panels.}
\label{f:astrometry_lofar_gmrt}
\end{figure*}

\subsection{Astrometry}\label{s:Astrometry}
To test the relative astrometry of the radio sources detected at different frequencies, we first cross-match the radio sources detected at LOFAR 150\,MHz, GMRT 610\,MHz, uGMRT 400\,MHz and 1,250\,MHz, and VLA 5\,GHz by using a matching radius of 10\,$\arcsec$. As shown in the first two panels of Figure~\ref{f:astrometry_lofar_gmrt}, the astrometry of GMRT 610\,MHz data (including GMRT-deep and GMRT-wide) is well-consistent with that of LOFAR 150\,MHz data with the median offsets of $\Delta$\,R.A.\,$\sim0\farcs1$ and the $\Delta$\,Decl.\,$\sim0\farcs04$. However, the median astrometric offsets between uGMRT 400\,MHz and LOFAR 150\,MHz data are $\sim0\farcs5$ in both right ascension and declination. Such offsets result in relatively larger cross-matching radii between 400\,MHz sources and the sources detected at LOFAR 150\,MHz, and GMRT 610\,MHz (Appendix~\ref{s:ugmrt_select}). The median astrometric offsets between uGMRT 1,250\,MHz and LOFAR 150\,MHz data are also relatively larger with $\Delta$\,R.A.\,$\sim0\farcs6$ and the $\Delta$\,Decl.\,$\sim0\farcs2$. The last panel of Figure~\ref{f:astrometry_lofar_gmrt} shows that the astrometry of VLA 5\,GHz data is well consistent with that of LOFAR 150\,MHz data with the median offsets of $\Delta$\,R.A.\,$=-0\farcs15$ and the $\Delta$\,Decl.\,$=0\farcs04$.

\subsection{Flux density calibration}\label{s:photometry}
Before measuring the radio spectrum of sources, we first investigate the flux density calibration of each radio dataset used in this work. We note that the GMRT 610\,MHz, uGMRT 400\,MHz and 1,250\,MHz, and VLA 5\,GHz observations used the same primary calibrators, namely 3C\,286 and 3C\,48, and the visibilities were calibrated, imaged and mosaicked by following the CASA standard procedures  \citep{Taylor14,Chakraborty19,Ocran20, Sinha23}. 
The LOFAR 150\,MHz data differ in the calibrator used (87GB 160333.2+573543) and in the techniques for calibration and imaging. However, a further flux scale calibration was performed by using the radio data available in the ELAIS-N1 field, including the GMRT 610\,MHz data. \cite{Sabater21} described the details of their flux scale calibration and reported a likely uncertainty of $\sim$6.5 per cent in the flux scale. We will discuss the effect of this uncertainty on our analyses in Section $\S$\ref{s:spectral indices}.

\subsection{SFGs selection}\label{s:SFGs_selection}
Except the VLA 5\,GHz data, which have a very limited sky coverage, the LOFAR 150\,MHz and GMRT 610\,MHz (deep) data are the two deepest datasets that are utilised in studying radio spectral properties of the sources in this work. We therefore directly use the cross-matched results as well as the classification of SFGs and AGN in the two radio catalogues as presented in \cite{Kondapally21}, \cite{Best23}, and \cite{Ocran20} in our analyses.

\subsubsection{Cross-matching and SFGs selection of 150\,MHz sources}\label{s:SFGs_selection_150}
For the 31,610 LOFAR 150\,MHz sources within the overlapping region, 97.6\% (30,839/31,610) of them have been identified with an optical, and/or NIR, and/or MIR counterpart in \cite{Kondapally21}. In our analyses, we use the photometric redshift and optical depth of these 150\,MHz source, which were estimated from the best-fit SEDs by using the aperture-matched UV to MIR photometries presented in \cite{Kondapally21}. The photometric redshift was estimated by using a hybrid method that combines template fitting and machine learning. The optical depth was defined as the natural logarithm of the difference between the unattenuated and attenuated best-fit model spectra at $V$-band. Details of these estimations are described in \cite{Duncan21}. 

To perform the source classification for the 150\,MHz sources, \cite{Best23} fitted the aperture-matched multi-wavelength photometries using four different SED fitting codes, two of which \citep[CIGALE and AGNfitter;][]{Noll09, Calistro16} included AGN templates in the fitting, and two of which \citep[MAGPHYS and BAGPIPES;][]{daCunha08, Carnall18} did not. In total, \cite{Best23} classified 3,289 sources as hosting optical/IR AGN activity. 

\cite{Best23} also classified 4,797 radio-excess AGN (of which 510 were also optical/IR selected AGN) by identifying sources that showed an excess of radio emission by more than 0.7\,dex compared to the expected SFR to radio luminosity relationship according to their ridgeline analysis. 1,314 sources remained unclassifiable \citep{Best23}. 

Combining the two sets of classifications, we obtain a sample of 22,720 SFGs, after removing all objects that are either classified as AGN through optical/IR or radio-excess selection, or are unclassifiable due to a lack of relevant data in one of the two diagnostics.

For the radio-detected SFGs, we also use the consensus stellar masses and SFRs from \cite{Best23} in Section $\S$\ref{s:results}. 

%
%
\begin{figure*}
\centering
\includegraphics[width=0.98\textwidth]{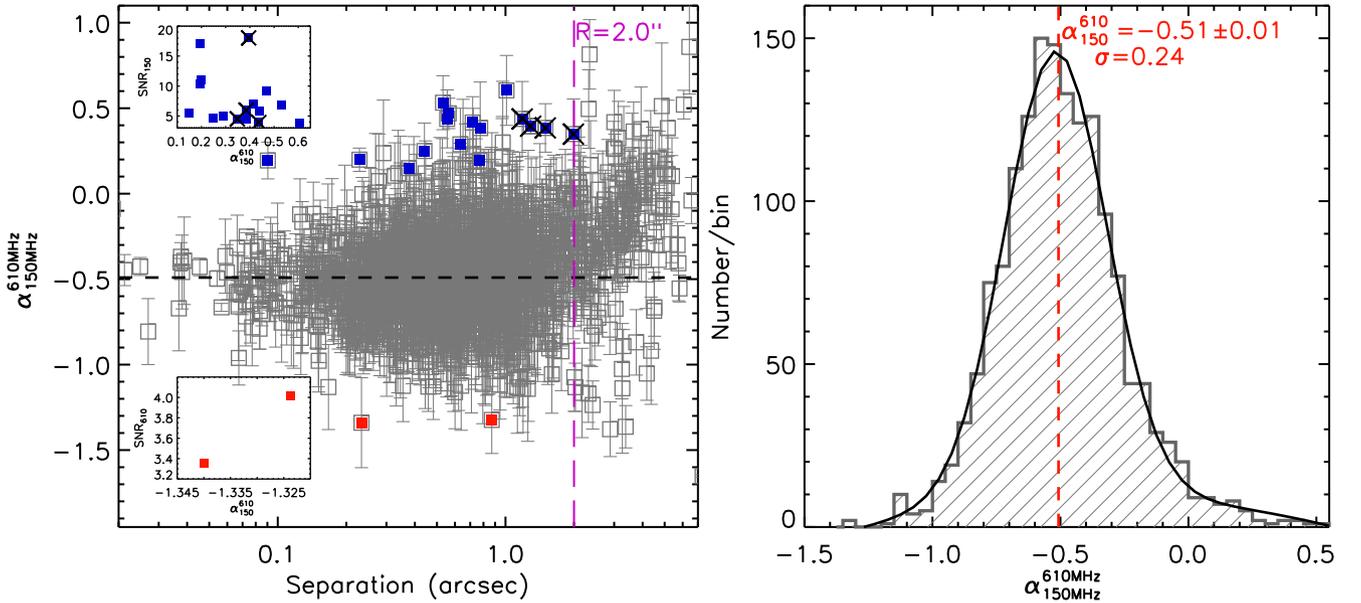}
\caption{{\it Left:} Radio spectral index between the observer-frame frequencies of 150--610\,MHz plotted against the separation between LOFAR 150\,MHz and GMRT 610\,MHz detections for the 1,782 SFGs. The blue squares are the 16 sources with $r<2\arcsec$ and $\alpha_{150}^{610}-\sigma>0.0$. We show their SNR at 150\,MHz as a function of $\alpha_{150}^{610}$ in the top inset plot. We visually inspect both LOFAR 150\,MHz and GMRT 610\,MHz images and find that four of these 14 sources are resolved as two components at LOFAR 150\,MHz, which are marked by black crosses. Therefore, we remove these four sources from our sample. There are also two sources with $\alpha_{150}^{610}+\sigma<-1.0$, which are marked by red squares. Visual inspection shows that both of them are faint at GMRT 610\,MHz. {\it Right:} Distribution of radio spectral index, $\alpha_{150}^{610}$, of 1,590 SFGs with $r<2\arcsec$. The median is $\alpha_{150}^{610}=-0.51\pm0.01$ with a scatter of $\sigma=0.24$.
}
\label{f:en1_gmrt_lofar_deep}
\end{figure*}

\subsubsection{Cross-matching and SFGs selection of 610\,MHz sources}\label{s:SFGs_610}
For the 4,290 GMRT 610\,MHz-detected sources in \cite{Ocran20}, 3,689 of them have a MIR counterpart from the SERVS and 3,542 have a NIR counterpart from the UKIDSS-DXS. Among them, 3,295 have been detected at both MIR and NIR. The multi-wavelength counterparts of these 610\,MHz sources were identified by using their MIR and/or NIR positions. Based on these cross-match results, \cite{Ocran20} carried out a multi-wavelength study using optical, X-ray, infrared, and radio diagnostics to identify AGN in their sample. We refer readers to \cite{Ocran20}  for more details of the cross-match and SFG/AGN classification. Here we briefly summarize the results of their classification. 

For the 4,290 GMRT 610\,MHz sources in \cite{Ocran20}, 3,490 of them have at least one measured multi-wavelength SFG/AGN classification diagnostic. Among them, 
\begin{itemize}
\item 209 are classified as MIR AGN by adopting the MIR colour-colour selection criteria in \cite{Donley12}, 
\item 106 are optical spectroscopic AGN by using the BOSS CLASS and SUBCLASS parameters \citep{Bolton12}, 
\item 102 are X-ray AGNs with $L_{\rm X}>10^{42}$\,erg\,s$^{-1}$, and 
\item 340 are classified as radio loud AGN by comparing their MIR (MIPS 24\,um) to radio flux ratio \citep{Bonzini13}. 
\end{itemize}
We therefore remove a total of 648 AGNs and obtain a sample of 2,842 SFGs from the GMRT 610\,MHz-detected sources. 

\subsubsection{Sample of SFGs}\label{s:SFGs}
To construct a final sample of SFGs for studying the radio spectral properties, we first compare the results of source classification from \cite{Ocran20} and \cite{Best23}. For the GMRT 610\,MHz SFGs that have a LOFAR 150\,MHz counterpart within 2$\arcsec$ (the cross-match radius of the two datasets discussed in Section $\S$\ref{s:gmrt_select}), 70\% of them are also classified as SFGs in \cite{Best23}. Meanwhile, 25\% of them are classified as AGN. Specifically, 17\% were classified as radio-excess AGN in \cite{Best23}. Therefore, the different selections of radio AGN cause the major difference between the source classifications in \cite{Ocran20} and \cite{Best23}. The remaining 5\% were unclassifiable due to a lack of measured diagnostic in \cite{Best23}. 

Due to the different sensitivities and coverages of the radio data used in this work, we include all SFGs classified in either \cite{Best23} or \cite{Ocran20} in the parent sample of SFGs. For the main results based on GMRT-deep-selected sources, we first use the sample of 2,842 SFGs classified in \cite{Ocran20} and perform a further AGN exclusion by removing sources that have a LOFAR 150\,MHz counterpart and are classified as AGN in \cite{Best23}. Besides, we also include the SFGs that are classified in \cite{Best23} but lack measured SFG/AGN classification diagnostic in \cite{Ocran20} in the final sample of SFGs (Section $\S$\ref{s:results}). For the results based on GMRT-wide or uGMRT 400\,MHz-detected sources, the SFG sample from \cite{Best23} is used as the parent sample, and a further AGN exclusive based on the classification from \cite{Ocran20} is performed. Details of the final SFG sample's composition are described in Sections $\S$\ref{s:results} and Appendix~\ref{s:gmrt_wide_select} and \ref{s:ugmrt_select}.

\section{Results} \label{s:results}
Because of the high sensitivity and large coverage of LOFAR 150\,MHz and GMRT 610\,MHz (deep) data, we first study the radio spectral properties of selected SFGs at observer-frame frequencies of 150--610\,MHz. The relatively higher sensitivity of LOFAR 150\,MHz data guarantees high completeness of 150\,MHz detection rate of 610\,MHz-selected sources (Section $\S$\ref{s:gmrt_select}). We therefore use 610\,MHz as the selection frequency in the following analyses. We also include the uGMRT 400\,MHz and 1,250\,MHz, and VLA 5\,GHz data in studying radio spectral properties at 150--5,000\,MHz if the SFGs are also detected at such frequencies as presented in Sections $\S$\ref{s:ugmrt_select_index}, $\S$\ref{s:uGMRT_1250MHz}, and $\S$\ref{s:610_5000}.

\subsection{Radio spectral index between the observer-frame frequencies of 150--610\,MHz}\label{s:gmrt_select}
To include all possible 150\,MHz counterparts of 610\,MHz-selected sources, we first use a cross-match radius of 10$\arcsec$. For the 4,290 GMRT 610\,MHz sources in \cite{Ocran20}, 87\% (3,748/4,290) of them have a LOFAR 150\,MHz counterpart within 10$\arcsec$. If we limit the GMRT 610\,MHz sources to those with an associated redshift in \cite{Ocran20}, this fraction increases to 90\% (2,782/3,105). For the 2,842 GMRT 610\,MHz-selected SFGs (Section $\S$\ref{s:SFGs_610}), 2,462 of them (87\%) have a LOFAR 150\,MHz counterpart within 10$\arcsec$.

We first measure the radio spectral index at observer-frame frequencies of 150--610\,MHz, $\alpha_{\rm 150}^{\rm 610}$, for a total of 2,462 GMRT 610\,MHz-selected SFGs, which have a LOFAR 150\,MHz counterpart within 10$\arcsec$. We assume a power-law radio spectrum between 150--610\,MHz and measure the two-point spectral index $\alpha_{150}^{610}$ by: 
\begin{eqnarray} \label{e:equation0}
\alpha=\frac{{\rm log} (S_{1}/S_{2})}{{\rm log} (\nu_{\rm 1}/\nu_{\rm 2})}, 
\end{eqnarray}
where $S_{1}$ and $S_{2}$ are the flux densities at frequencies $\nu_{1}$ and $\nu_{2}$ respectively. The effective frequency ($\nu$) of each radio dataset used in this work is listed in Table \ref{tab:table1}. The uncertainty of $\alpha_{\rm 150}^{\rm 610}$ ($\sigma_{150}^{610}$) is inherited from the uncertainties of the two flux densities. We show the $\alpha_{\rm 150}^{\rm 610}$ as a function of the separation between the GMRT 610\,MHz and LOFAR 150\,MHz detections in Figure~\ref{f:en1_gmrt_lofar_deep}. The distribution of $\alpha_{\rm 150}^{\rm 610}$ becomes asymmetric when the separation is greater than $2\arcsec$ as shown in Figure~\ref{f:en1_gmrt_lofar_deep}. We therefore choose $r_{\rm 150-610}=2\arcsec$ as the cross-match radius when identifying the LOFAR 150\,MHz counterparts for the 610\,MHz-selected SFGs. To reduce the contamination from AGN, we remove additional 590 GMRT 610\,MHz sources that have a LOFAR 150\,MHz counterpart within 2$\arcsec$ and are classified as AGN in \cite{Best23}. 

Among the remaining 1,782 SFGs, 16 of them have the measured $\alpha_{150}^{610}-\sigma_{150}^{610}>0.0$ (highly flat radio spectrum at 150--610\,MHz). We show their SNR at 150\,MHz as a function of $\alpha_{150}^{610}$ in the top-left inset plot in Figure~\ref{f:en1_gmrt_lofar_deep}. We find that those SFGs with a flatter radio spectrum do have a lower SNR at 150\,MHz. We also visually inspect both LOFAR 150\,MHz and GMRT 610\,MHz images and find that four of them are resolved as two components at LOFAR 150\,MHz. 
We, therefore, remove them from our sample. There are also two sources with $\alpha_{150}^{610}+\sigma_{150}^{610}<-1.0$ (highly steep radio spectrum at 150--610\,MHz). We inspect the LOFAR 150\,MHz and GMRT 610\,MHz images and find that these two sources are well-matched but are relatively faint at GMRT 610\,MHz.

There are 1,590 SFGs with $r_{\rm 150-610}<2\arcsec$ after we remove the four SFGs that are resolved at 150\,MHz. We show the distribution of their $\alpha_{\rm 150}^{\rm 610}$ in the right plot of Figure~\ref{f:en1_gmrt_lofar_deep}. The median is $\alpha_{150}^{ 610}=-0.51\pm0.01$ with a scatter of $\sigma=0.24$ for these SFGs. The error of the median $\alpha_{150}^{ 610}$ is estimated by bootstrap resampling.

%
%
\begin{figure}
\centering
\includegraphics[width=0.48\textwidth]{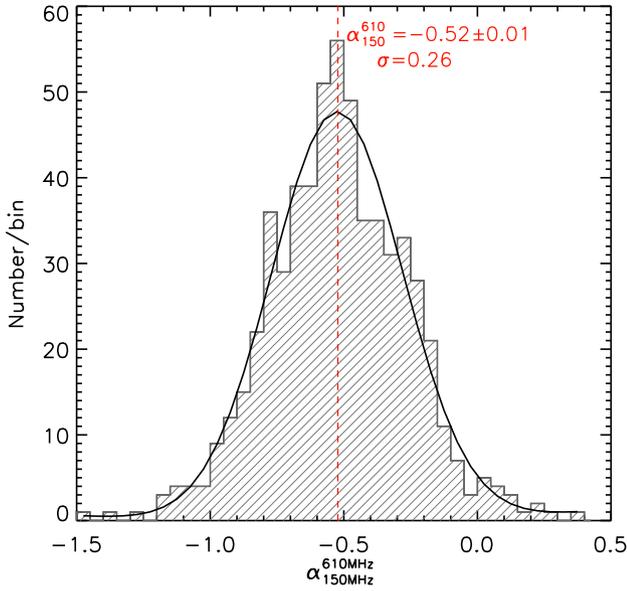}
\caption{Distribution of radio spectral index at observed 150--610\,MHz for the 596 GMRT 610\,MHz-detected sources that lack measured SFG/AGN classification diagnostic in the GMRT-deep catalogue \citep{Ocran20}. These sources have LOFAR 150\,MHz counterparts within 2$\farcs$0 and are classified as SFGs in the LOFAR multi-wavelength catalogue \citep{Best23}. The median is $\alpha_{150}^{610}=-0.52\pm0.01$ (red dashed line) with a scatter of 0.26, which is consistent with the results from the sources classified as the SFGs in the GMRT-deep catalogue \citep{Ocran20}, i.e., the results shown in Figure~\ref{f:en1_gmrt_lofar_deep}.}
\label{f:en1_gmrt_lofar_deep_hist_wrl}
\end{figure}

For the 4,290 GMRT 610\,MHz-detected sources, 800 of them lack a measured diagnostic to classify them as SFGs or AGN in \cite{Ocran20}. Among them, 667 have a LOFAR 150\,MHz counterparts within $2\arcsec$, and 596/667 are classified as SFGs in \cite{Best23}. As shown in Figure~\ref{f:en1_gmrt_lofar_deep_hist_wrl}, the median $\alpha_{\rm 150}^{\rm 610}$ of these 596 SFGs is $\alpha_{\rm 150}^{\rm 610}=-0.52\pm0.01$ with a standard deviation of $\sigma=0.26$, which is highly consistent with the results shown in Figure~\ref{f:en1_gmrt_lofar_deep}.

Therefore, the final sample used to study radio spectral properties at 150--610\,MHz consists of the 1,590 SFGs that are classified in \cite{Ocran20} and the 596 SFGs that are classified in \cite{Best23}, totalling 2,186 SFGs.

For the GMRT 610\,MHz-selected sources in \cite{Ocran20}, although the completeness of the detections at LOFAR 150\,MHz is $\sim$90\%, we further check these sources without LOFAR 150\,MHz counterparts. There are 327 GMRT 610\,MHz-selected sources with estimated photometric redshift in \cite{Ocran20} but not included in the LOFAR final cross-matched catalogue \citep{Kondapally21}. Our visual inspection shows that most of them are within the poor-quality areas of the LOFAR map, such as near a bright source. We directly measure the pixel value at their 610\,MHz positions and find 92\% of them have the value $>100\,\mu$Jy\,beam$^{-1}$, which would correspond to a $>\,5\sigma$ detection in a cleaner part of the LOFAR image. Therefore, although these sources are not included in the LOFAR catalogue because of the inability to fit with a Gaussian profile, there is no reason to expect that this will bias our results.

%
%
\begin{figure*}
\centering
\includegraphics[width=0.98\textwidth]{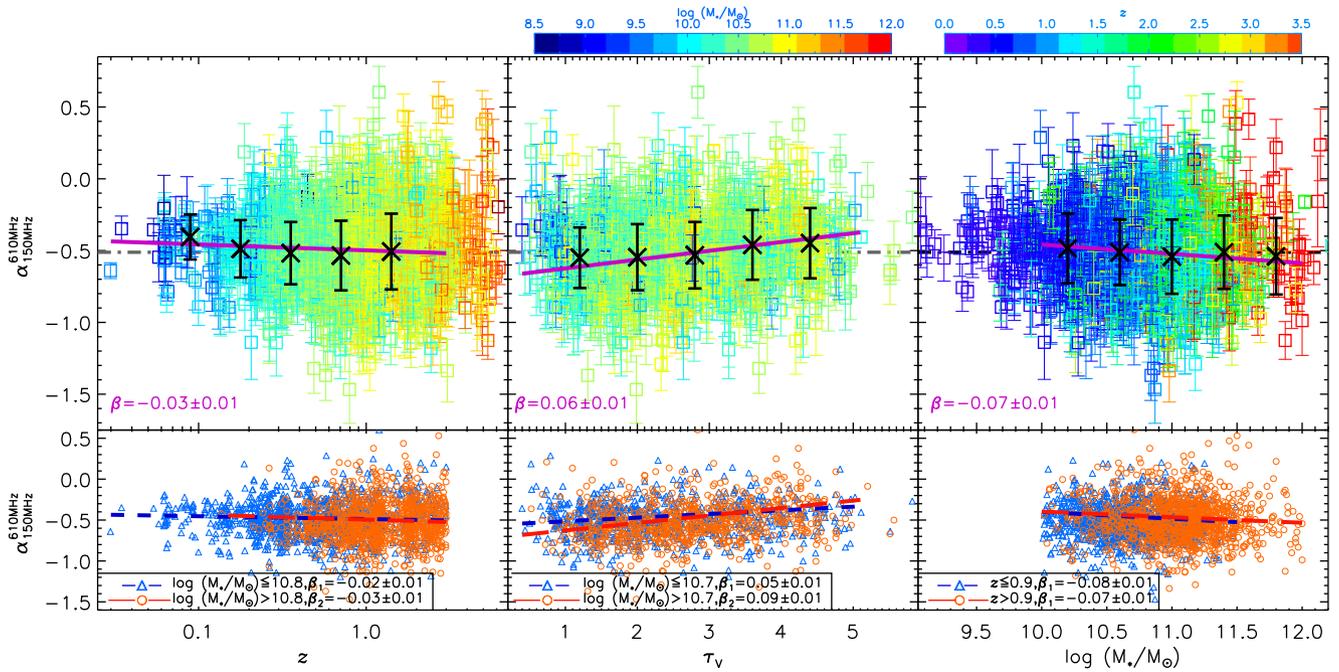}
\caption{Radio spectral index between observer-frame frequencies of 150--610\,MHz ($\alpha_{150}^{610}$) as functions of redshift ({\it left}), optical depth at $V$-band ({\it middle}), and stellar mass ({\it right}) for the GMRT 610\,MHz-selected SFGs. We have estimated photometric redshift and stellar mass for all of the 2,186 SFGs, and optical depth at $V$-band, $\tau_{V}$, for the 1,481 SFG with $z<1.5$. The black cross represents the median $\alpha_{\rm 150}^{610}$ of SFGs in each bin while the error bar shows the standard deviation.
The colour of symbols represents the stellar mass in the left-top and middle-top panels and represents the redshift in the right-top panel as shown by the two color bars. The linear fits are performed by using the individual measurements of SFGs and the results are shown by the purple solid lines. We find that, on average, the radio spectral slope of SFGs at observed 150--610\,MHz slightly flattens at low redshift (low frequency), flattens with increasing optical depth, and steepens with increasing stellar mass. To reduce the effect from the incompleteness of our radio flux-limited sample, we limit SFGs to $z<3$ and log $(M_{*}/\Msun)>10$ when fitting the correlations between $\alpha_{150}^{610}$ and redshift, and stellar mass respectively.
{\it Bottom:} We further divide our sample into the massive and the less-massive subsamples and find that the correlations between $\alpha_{150}^{610}$ and redshift, optical depth at $V$-band are independent from the stellar mass. When studying the correlation between $\alpha_{150}^{610}$ and stellar mass, we divide our sample into the high-redshift ($z>0.9$, 1,049) and the low-redshift ($z\le0.9$, 1,015) subsamples and find that this correlation is independent from redshift.}
\label{f:en1_gmrt_lofar_deep_matched_alpha_z_Ms_final_conc}
\end{figure*}

\subsubsection{Radio spectral properties at 150--610\,MHz}\label{s:gmrt_select_spectral_properties}
For the GMRT 610\,MHz-selected SFGs, we show their radio spectral index between observer-frame frequencies of 150--610\,MHz as functions of redshift, optical depth, and stellar mass in Figure~\ref{f:en1_gmrt_lofar_deep_matched_alpha_z_Ms_final_conc}. The photometric redshifts for the total 2,186 SFGs are from the LOFAR multi-wavelength photometric catalogue \citep{Duncan21}. To reduce the effect of the incompleteness of our radio flux-limited sample, we limit SFGs to $z<3$ when fitting the correlations between $\alpha_{150}^{610}$ and  redshift. We perform a linear fit by minimizing the $\chi^{2}$ and using the uncertainties of $\alpha_{150}^{610}$ as the inverse weights. As shown in the left panel of Figure~\ref{f:en1_gmrt_lofar_deep_matched_alpha_z_Ms_final_conc}, on average, the radio spectral slope at observer-frame 150--610\,MHz slightly flattens at low redshift (low frequency) with a $\sim$3\,$\sigma$ significance. We further divide our sample into the massive (log (M$_{*}/\Msun)>10.8$) and less-massive (log (M$_{*}/\Msun)\le10.8$) subsamples with sample sizes of 1,040 and 1,020 respectively. The correlation between $\alpha_{150}^{610}$ and redshift of the two subsamples are consistent with that of the full sample, although the less-massive subsample is affected more by the incompleteness of our radio flux-limited sample.

In our sample, there are 1,481 SFGs with $z<1.5$, which have estimated optical depth, $\tau_{V}$, from \cite{Smith21}. As shown in the middle panel of Figure~\ref{f:en1_gmrt_lofar_deep_matched_alpha_z_Ms_final_conc}, the radio spectrum at observer-frame 150--610\,MHz flattens with increasing optical depth. The linear fit slope of this correlation is $\beta=$ 0.06$\pm0.01$. We also divide the sample into massive (log (M$_{*}/\Msun)>10.7$) and less-massive (log (M$_{*}/\Msun)\le10.7$) subsets with sample sizes of 718 and 763 respectively. We obtain the linear fit slopes of $\beta=0.09\pm0.01$ and $\beta=0.05\pm0.01$ for the two subsets respectively, which confirms that this trend is independent from the stellar mass of SFGs. 

In our radio flux-limited sample, the low-mass galaxies become increasingly under-represented with increasing redshift (Figure~\ref{f:en1_gmrt_lofar_deep_matched_alpha_z_Ms_final_conc}). We therefore limit our sample to log (M$_{*}/\Msun)>10$ when fitting the correlations between $\alpha_{150}^{610}$ and stellar mass as shown in the right panel of Figure~\ref{f:en1_gmrt_lofar_deep_matched_alpha_z_Ms_final_conc}. On average, the radio spectral slope at observer-frame 150--610\,MHz slightly steepens with increasing stellar mass for these radio-selected SFGs with a linear fit slope of $\beta=-0.07\pm0.01$. We further divide our sample into high-redshift ($z>0.9$) and low-redshift ($z\le0.9$) subsets and find that this trend is independent from redshift, with the linear fit slopes being $\beta=-0.11\pm0.01$ and $\beta=-0.08\pm0.01$ respectively. These results are consistent with our previous work based on the MeerKAT and VLA data in the COSMOS field \citep{An21}, although for a relatively higher frequency (observer-frame frequencies of 1.3--3\,GHz).

We also use GMRT-wide as the selection frequency and obtain the results that are well-consistent with that based on GMRT-deep data (Appendix~\ref{s:gmrt_wide_select}).

\subsection{Radio spectral properties at observed 150--400\,MHz and 400--610\,MHz}\label{s:ugmrt_select_index}
We include the uGMRT 400\,MHz data \citep{Chakraborty19} in studying the radio spectrum of SFGs between observer-frame frequencies of 150--400\,MHz and 400--610\,MHz respectively. For the 2,186 SFGs that have both GMRT-deep 610\,MHz and LOFAR 150\,MHz detections, 1,043 of them have a 400\,MHz counterpart within 2$\farcs$5 (the cross-matching radius of GMRT-deep 610\,MHz and uGMRT 400\,MHz data as described in Appendix~\ref{s:ugmrt_select}). Their median two-point spectral indices are $\alpha_{150}^{400}=-0.30\pm0.01$, $\alpha_{400}^{610}=-0.98^{+0.03}_{-0.02}$, and $\alpha_{150}^{610}=-0.51\pm0.01$ respectively. Therefore, the radio spectral indices we measured are flatter over the 150--400\,MHz and steeper over the 400--610\,MHz. As we discussed in Section $\S$\ref{s:spectral indices} and Appendix~\ref{s:ugmrt_select}, this is caused by the relatively shallower uGMRT 400\,MHz data, namely the Eddington bias.

To get a less biased sample for studying radio spectrum at observer frame frequencies of 150--400\,MHz and 400--610\,MHz, we instead apply a flux density cut at GMRT 610\,MHz. The coverages of GMRT-deep 610\,MHz and uGMRT 400\,MHz are slightly offset as shown in Figure~\ref{f:coverage}. Within the overlapped region, only 47\% (2,024/4,269) of GMRT 610\,MHz sources have a 400\,MHz counterpart within 10$\arcsec$. We therefore limit the sources to $S_{\rm 610\,MHz}>300\,\mu$Jy to guarantee that $>$80\% of GMRT 610\,MHz sources have a 400\,MHz counterpart within 10$\arcsec$. Within the overlapped region, there are 303 SFGs with $S_{\rm 610\,MHz}>300\,\mu$Jy. Among them, 258 have a 400\,MHz counterpart within 2$\farcs$5. All of them also have a LOFAR 150\,MHz counterpart within 2$\arcsec$ thanks to the high sensitivity of LOFAR data. The median spectral indices between observer-frame frequencies of 150--400\,MHz, 400--610\,MHz, and 150--610\,MHz are $\alpha_{150}^{400}=-0.40\pm0.02$, $\alpha_{400}^{610}=-0.49^{+0.05}_{-0.04}$, and $\alpha_{150}^{610}=-0.42^{+0.02}_{-0.01}$. 

\begin{figure*}
\centering
\includegraphics[width=0.98\textwidth]{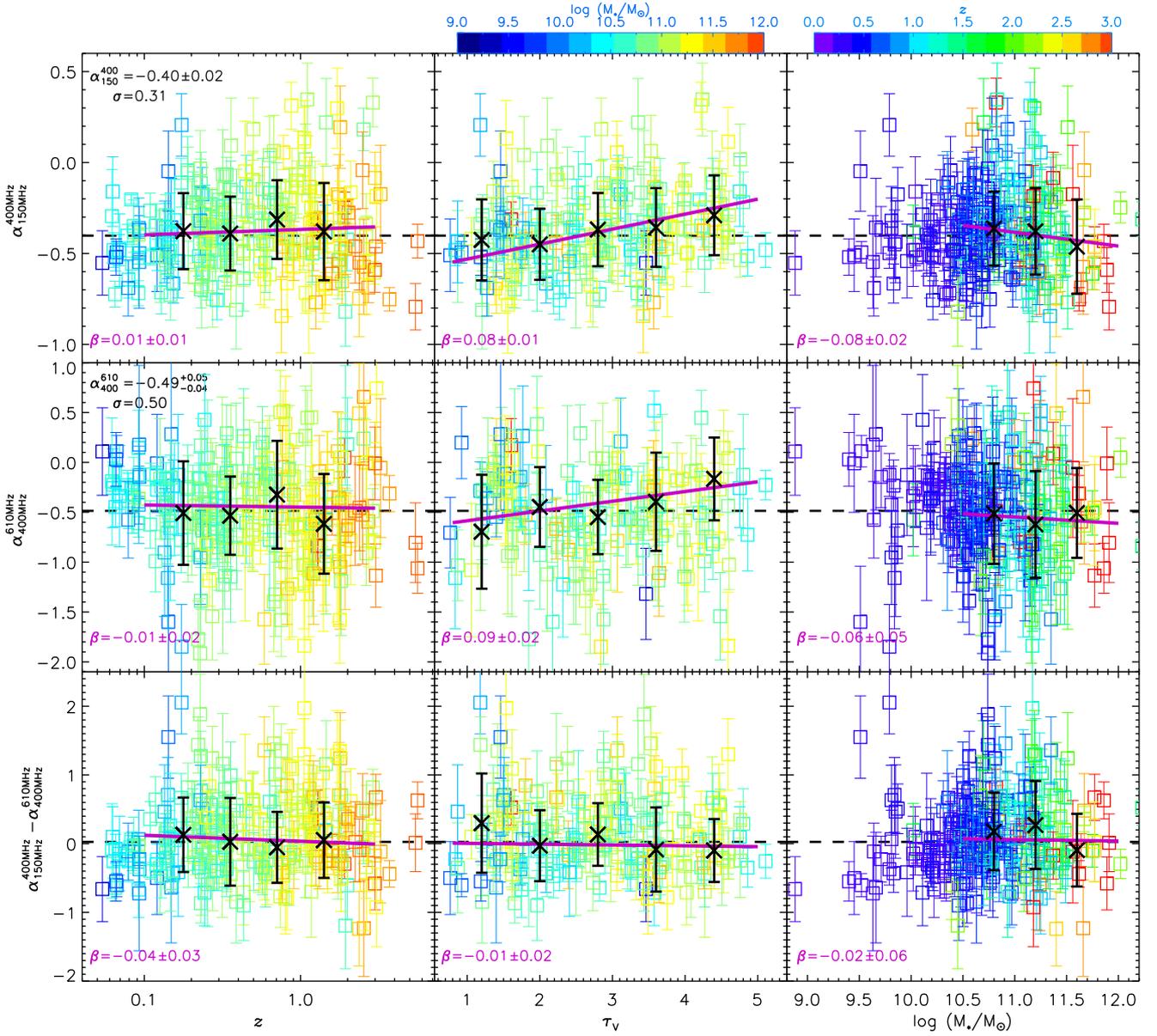}
\caption{Radio spectral indices at observer-frame frequencies of 150--400\,MHz (first row), 400--610\,MHz (second row), and the difference between them (third row) as functions of redshift (left column), optical depth at $V$-band (middle column), and stellar mass (right column) for the 258 SFGs that have $S_{\rm 610\,MHz}>300\,\mu$Jy, and detections at both uGMRT 400\,MHz and LOFAR 150\,MHz. The black crosses are the median $\alpha_{150}^{400}$, $\alpha_{400}^{610}$, and ($\alpha_{150}^{400}-\alpha_{400}^{610}$) of SFGs within each bin while the error bars show the scatters. The purple solid lines show the linear fit results, which are based on the individual measurements of SFGs. As shown in the first column, the radio spectral index at 150--400\,MHz is not significantly correlated with redshift, i.e., rest-frame frequency. The second and third columns show that, on average, the radio spectra at both frequency ranges slightly flatten with optical depth and steepen with stellar mass, although the uncertainty is large, especially for the correlation between $\alpha_{400}^{610}$ and stellar mass. However, the difference between $\alpha_{150}^{400}$ and $\alpha_{400}^{610}$ does not have significant correlations with these physical properties.}
\label{f:en1_gmrt_deep_lofar_ugmrt_matched_alpha_z_Ms_final_curve_color_1}
\end{figure*}

We show the correlations between the radio spectral indices and redshift, optical depth at $V$-band, and stellar mass of these 258 SFGs in Figure~\ref{f:en1_gmrt_deep_lofar_ugmrt_matched_alpha_z_Ms_final_curve_color_1}. To reduce the effect of the incompleteness of our radio flux-limited
sample, we limit SFGs to $z<3$ and log (M$_{*}/\Msun)>10.5$ when fitting the correlations between radio spectral indices and redshift, and stellar mass respectively. As shown in Figure~\ref{f:en1_gmrt_deep_lofar_ugmrt_matched_alpha_z_Ms_final_curve_color_1}, at $z<0.1$, our sample is dominated by the SFGs with $\alpha_{400}^{610}>-0.50$, which is very likely a selection bias caused by the flux density cut at 610\,MHz, i.e., $S_{\rm 610\,MHz}>300\,\mu$Jy. 
We, therefore, use SFGs with $0.1<z<3$ to fit correlations between radio spectral indices and redshift (Figure~\ref{f:en1_gmrt_deep_lofar_ugmrt_matched_alpha_z_Ms_final_curve_color_1}). We find that, on average, the radio spectra at observer-frame frequencies of 150--400\,MHz and 400--610\,MHz are not significantly correlated with redshift, i.e., rest-frame frequency, but flatten with increasing optical depth and steepen with increasing stellar mass. Consequently, the difference between radio spectral indices at the observer-frame frequencies of 150--400\,MHz and 400--610\,MHz is not correlated with these physical properties as shown in the third row of Figure~\ref{f:en1_gmrt_deep_lofar_ugmrt_matched_alpha_z_Ms_final_curve_color_1}. 

%
%

\subsection{Radio spectral properties at 150--1,250\,MHz}\label{s:uGMRT_1250MHz}
We also cross-match the uGMRT 1,250\,MHz sources detected in \cite{Sinha23} with GMRT-deep 610\,MHz sources by using a radius of 1$\farcs$5. Details of choosing the cross-matching radius are presented in Appendix~\ref{s:1250_select}. Within the uGMRT 1,250\,MHz source detection region, only $\sim$40\% of GMRT-deep 610\,MHz sources have a 1,250\,MHz counterpart because of the shallow uGMRT 1,250\,MHz data. We include all of the 1,086 sources detected in \cite{Sinha23} in our analyses but point out that the following results are only for SFGs brighter than 60\,$\mu$Jy at 1,250\,MHz. 

For SFGs with both GMRT-deep 610\,MHz and LOFAR 150\,MHz detections, 324 of them have an uGMRT 1,250\,MHz counterpart within 1$\farcs$5. Furthermore, we remove nine sources that are resolved at 1,250\,MHz. For the remaining 315 SFGs, we obtain the median two-point spectral indices of $\alpha_{150}^{1250}=-0.48\pm0.01$ and $\alpha_{610}^{1250}=-0.56^{+0.04}_{-0.02}$ with a scatter of $\sigma=$\,0.20 and 0.43 respectively. These are 284 SFGs that also have uGMRT 400\,MHz detections. We obtain a median spectral index of $\alpha_{400}^{1250}=-0.68\pm0.04$ and a scatter of $\sigma=0.51$.

\begin{figure*}
\centering
\includegraphics[width=0.98\textwidth]{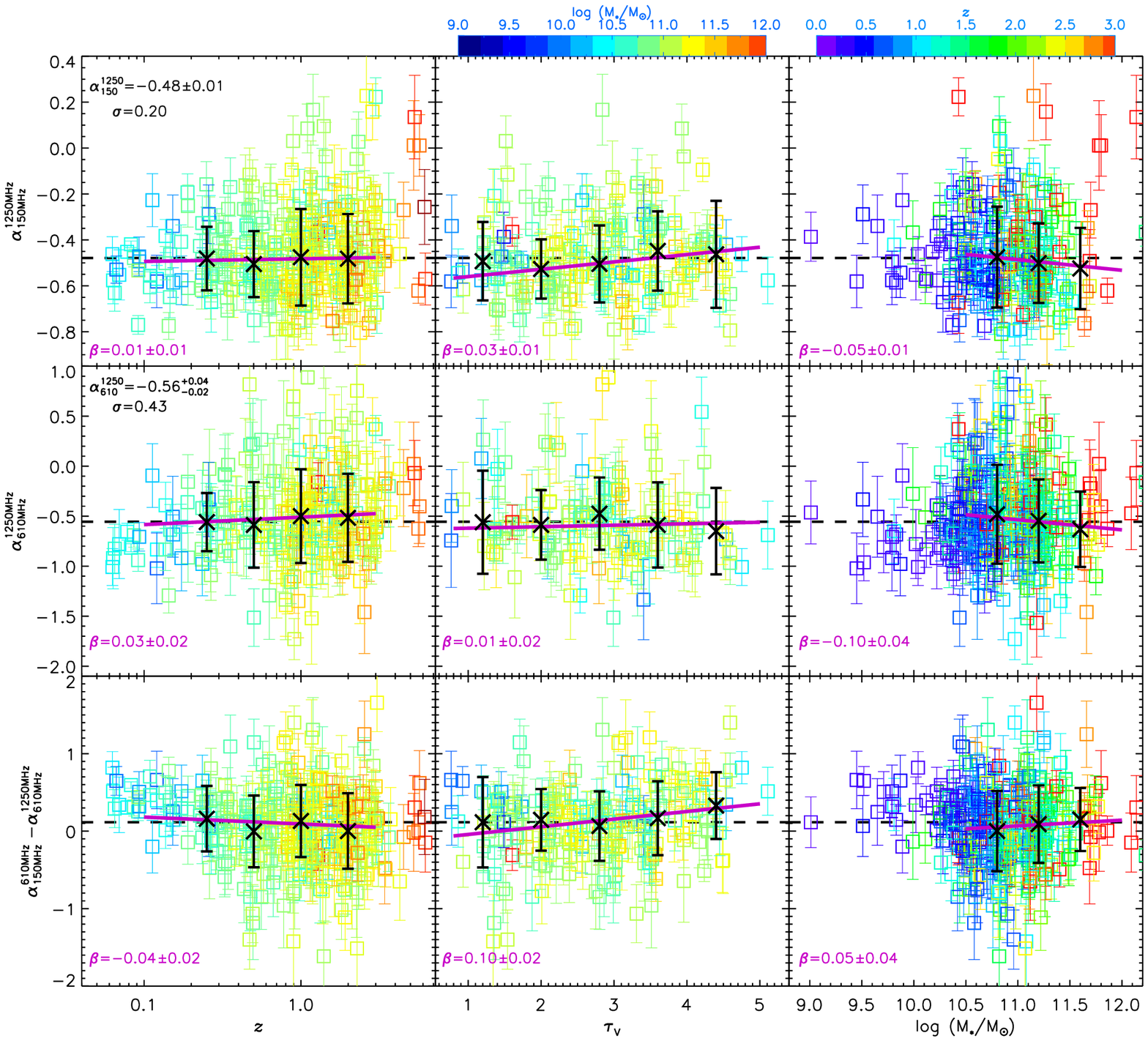}
\caption{Radio spectral indices at observer-frame 150--1,250\,MHz (first row), 610--1,250\,MHz (second row), and the spectral curvature at 610\,MHz ($\alpha_{150}^{610}-\alpha_{610}^{1250}$, last row) as functions of redshift, optical depth at $V$-band ($\tau_{V}$), and stellar mass of SFGs respectively. The first column shows that the radio spectral index at 150--1,250\,MHz is not significantly correlated with redshift, i.e., rest-frame frequency. The second column shows that there is a weak positive correlation between $\alpha_{150}^{1250}$ and $\tau_{V}$ but this correlation is negligible at observed 610--1,250\,MHz. The bottom panel of the second column shows that the difference between radio spectra indices at observer-frame 150--610\,MHz and 610--1,250\,MHz positively correlated with $\tau_{V}$.
The last column shows that the radio spectrum slightly steepens with increasing stellar mass at observed 150--1,250\,MHz.}
\label{f:en1_gmrt_deep_1p2}
\end{figure*}

We correlate the radio spectral indices at 150--1,250\,MHz and 610--1,250\,MHz with redshift, optical depth at $V$-band, and stellar mass for the 315 SFGs that have detections at these three frequencies. To compare the results at low frequencies and reduce the effect of the incompleteness of our radio flux-limited sample, we also limit SFGs to $0.1<z<3$ and log (M$_{*}/\Msun)>10.5$ when correlating with redshift and with stellar mass respectively. Similarly to the results shown in Section~$\S$\ref{s:ugmrt_select_index}, we find that the correlation between radio spectral index and redshift is negligible and the radio spectrum slightly steepens with increasing stellar mass at observed 150--1,250\,MHz, although the scatter is large. However, the correlation between radio spectral index and optical depth at $V$-band is weaker at 150--1,250\,MHz and even negligible at observed 610--1,250\,MHz as shown in Figure~\ref{f:en1_gmrt_deep_1p2}, which is in contrast to our findings at low frequencies ($\nu\la$1\,GHz, Figure~\ref{f:en1_gmrt_lofar_deep_matched_alpha_z_Ms_final_conc} and Figure~\ref{f:en1_gmrt_deep_lofar_ugmrt_matched_alpha_z_Ms_final_curve_color_1}). This leads to a positive correlation between the spectral curvature, i.e., ($\alpha_{150}^{610}-\alpha_{610}^{1250}$), and optical depth at $V$-band as shown in Figure~\ref{f:en1_gmrt_deep_1p2}.

%
%

%
%
\begin{figure}
\centering
\includegraphics[width=0.48\textwidth]{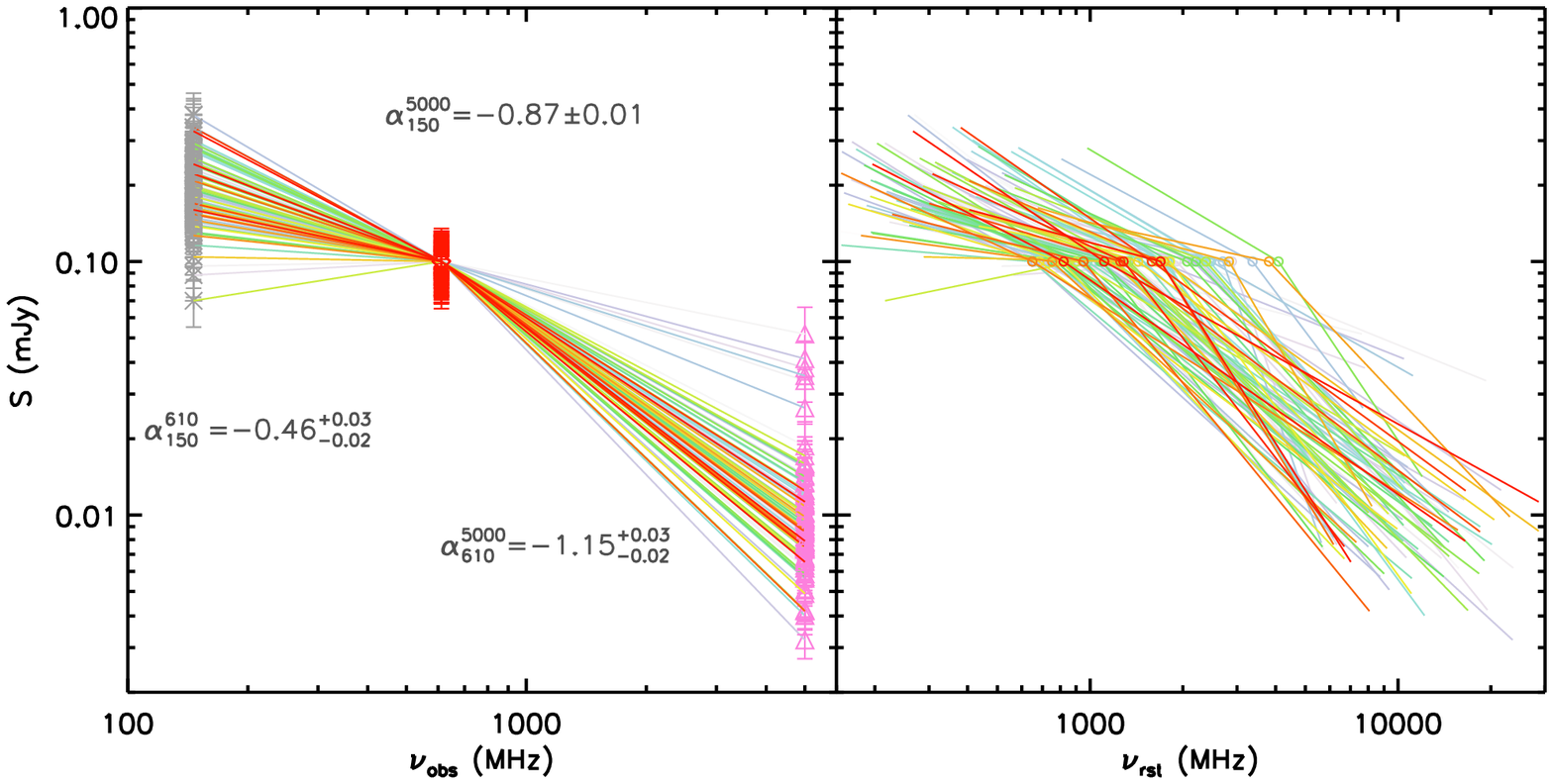}
\caption{Observer-frame ({\it left}) and rest-frame ({\it right}) radio spectra of 94 SFGs that have detections at 150\,MHz, 610\,MHz, and 5\,GHz. Their 610\,MHz flux densities are normalized to 0.1\,mJy for comparison. The median radio spectral indices between the observer-frame frequencies of 150--610\,MHz, 610--5,000\,MHz, and 150--5,000\,MHz are $\alpha_{150}^{610}=-0.46^{+0.03}_{-0.02}$, $\alpha_{610}^{5000}=-1.15^{+0.03}_{-0.02}$, and $\alpha_{150}^{5000}=-0.87\pm0.01$. For these 94 SFGs, 82\% (77/94) have a low-frequency radio spectrum significantly flatter than the high-frequency spectrum.}
\label{f:en1_gmrt_lofar_vla_SED_nml_restframe}
\end{figure}

\subsection{Radio spectral properties at 150--5,000\,MHz}\label{s:610_5000}

Within the VLA 5\,GHz coverage (the central 0.13\,deg$^{2}$ of the ELAIS-N1 field, Figure~\ref{f:coverage}), there are 121 GMRT 610\,MHz-detected SFGs. Among them, 100 have a VLA 5 GHz counterpart within 3$\arcsec$. We use a cross-matching radius of 3$\arcsec$ because of the relatively higher angular resolution of the VLA 5\,GHz data (Section~$\S$\ref{s:observation}). We find that applying a smaller matching radius, i.e., 1$\farcs$5, only makes the median radio spectrum at 610-5,000\,MHz slightly flatter. We therefore keep all of the VLA 5\,GHz sources that are matched with the 610\,MHz SFGs within 3$\arcsec$ and obtain a median spectral index of $\alpha_{610}^{5000}=-1.14^{+0.03}_{-0.04}$ with a scatter of $\sigma=0.27$.

 Among these 100 SFGs, 94 also have LOFAR 150\,MHz detection. The median two-point radio spectral indices between the observer-frame frequencies of 150--610\,MHz, 610--5,000\,MHz, and 150--5,000\,MHz are $\alpha_{150}^{610}=-0.46^{+0.03}_{-0.02}$, $\alpha_{610}^{5000}=-1.15^{+0.03}_{-0.02}$, and $\alpha_{150}^{5000}=-0.87\pm0.01$ with the scatters of $\sigma=$\,0.22, 0.23 and 0.15 respectively. Although the sample size is very limited and the VLA 5\,GHz data is deeper than the data at the other two frequencies, we find that 82\% (77/94) of these SFGs have their low-frequency radio spectrum significantly flatter than that at high frequency, namely, $\alpha_{150}^{610}-\sigma_{150}^{610}>\alpha_{610}^{5000}+\sigma_{610}^{5000}$  (Figure~\ref{f:en1_gmrt_lofar_vla_SED_nml_restframe}).

There are 65/94 and 47/94 SFGs, that also have uGMRT 400\,MHz and 1,250\,MHz detections respectively. However, because of the relatively shallower uGMRT 400\,MHz and 1,250\,MHz data, we only use LOFAR 150\,MHz, GMRT 610\,MHz, and VLA 5\,GHz data in studying radio spectral properties at 150--5,000\,MHz. We find that, at observer-frame frequencies of 150--5,000\,MHz, the radio spectrum of SFGs slightly steepens with increasing stellar mass, but there is no significant correlation between radio spectral index and optical depth, which is consistent with the results we find at observed 610--1,250\,MHz (Section$\S$\ref{s:uGMRT_1250MHz}).

\section{Discussion}\label{s:discussion}
By combining the LOFAR 150\,MHz, GMRT 610\,MHz, uGMRT 400\,MHz and 1,250\,MHz, and VLA 5\,GHz data in the ELAIS-N1 field, we study the radio spectral properties at observer-frame frequencies of 150--5,000\,MHz. Here we discuss the selection bias of this work and possible physical mechanisms that determine the radio spectrum at 150--5,000\,MHz. 

\subsection{Radio spectral indices at observer-frame 150--5,000\,MHz and selection bias}\label{s:spectral indices}
By using the two radio datasets with the greatest sensitivity and coverage in this work, we obtain a total of $\sim$3,500 SFGs that have detections at both GMRT 610\,MHz (including both GMRT-deep and GMRT-wide) and LOFAR 150\,MHz and a median spectral index of $\alpha_{150}^{610}=-0.51\pm0.01$. The photometric redshift of these SFGs ranges from $z=0.01$ to $z=6.21$, although the redshift uncertainty is large at the higher end \citep{Duncan21}. Our measurement is consistent with the median spectral indices at similar frequency ranges of nearby galaxies \citep{Marvil15, Chyzy18, Klein18, Heesen22}, submillimeter-selected galaxies \citep[SMGs,][]{Ramasawmy21}, and uGMRT 400\,MHz-detected sources \citep{Sinha23}. We note that \cite{Chyzy18}, \cite{Ramasawmy21}, and \cite{Sinha23} also include the $\sim$150\,MHz data in their analyses from the LOFAR Multifrequency Snapshot Sky Survey and the LoTSS. In addition, \cite{Sinha23} measured radio spectral indices for all uGMRT 400\,MHz sources, and the measurements in \cite{Ramasawmy21} are for submillimeter-brighter sources, which also include a small fraction of MIR color-color-selected AGN. For our work, including AGN results in a median spectral index of $\alpha_{150}^{610}=-0.49\pm0.01$. Therefore, the median radio spectrum is slightly flatter but consistent with that of SFGs within uncertainties.

As described in Section $\S$\ref{s:photometry}, the flux calibration of LOFAR 150\,MHz data has a likely uncertainty of $\sim$6.5 per cent in the flux scale. We include this uncertainty and obtain a median two-point radio spectral index of $\alpha_{150}^{610}=-0.51\pm0.05$ for GMRT-deep 610\,MHz-detected SFGs. Therefore, compared with the median radio spectral index at high frequency, specifically, at rest-frame $\sim$1.3--10\,GHz in \cite{An21}, as well as the synchrotron spectrum, the radio spectrum of SFGs is flatter at $\nu_{\rm rest}\la$1\,GHz. 

For GMRT-deep 610\,MHz-detected SFGs, we also include uGMRT 400\,MHz and 1,250\,MHz, and VLA 5\,GHz data to measure the radio spectral indices at 150--400--610--1,250\,MHz and 150--610--5,000\,MHz respectively. The minimum RMS noise  of GMRT-deep 610\,MHz data is 7.1\,$\mu$Jy beam$^{-1}$, which corresponds to $\sim$9\,$\mu$Jy beam$^{-1}$ at 400\,MHz, and $\sim$5\,$\mu$Jy beam$^{-1}$ at 1,250\,MHz if we adopt the median radio spectral index at 150--610\,MHz, i.e., $\alpha_{150}^{610}=-0.51$. The uGMRT 400\,MHz data used in this work are shallower than that, which biases our sample to SFGs with a relatively steeper radio spectrum at 400--610\,MHz and to SFGs with a relatively flatter radio spectrum at 150--400\,MHz (the Eddington bias) when we measuring the radio spectral indices at observed 150--400--610\,MHz as reported in Section $\S$\ref{s:ugmrt_select_index}. To get the less-biased results, we limit the sample to SFGs with $S_{\rm 610\,MHz}>300\,\mu$Jy in studying radio spectral properties at 150--400--610\,MHz (Section $\S$\ref{s:ugmrt_select_index}).

The radio spectrum at 610--1,250\,MHz is expected to be steeper than that at 150--610\,MHz because of the effects we discussed in Section $\S$\ref{s:flatter at low frequency}. Therefore, the corresponding sensitivity of the GMRT-deep 610\,MHz data is expected to be deeper than $\sim$5\,$\mu$Jy beam$^{-1}$ at 1,250\,MH. The shallow of the uGMRT 1,250\,MHz data biases our sample to SFGs with a relatively flatter radio spectrum at 610--1,250\,MHz and also at 150--1,250\,MHz. We, therefore, noted in Section $\S$\ref{s:uGMRT_1250MHz} that the radio spectral properties we obtained at 150--1,250\,MHz are only for SFGs that are brighter than 60$\,\mu$Jy at 1,250\,MHz.

VLA 5\,GHz data have a median RMS noise of 1.05\,$\mu$Jy, which corresponds to $\sim$3$\,\mu$Jy, $\sim$6$\,\mu$Jy, $\sim$8\,$\mu$Jy, and $\sim$17\,$\mu$Jy at 1,250\,MHz, 610\,MHz, 400\,MHz, and 150\,MHz respectively if we assume a radio spectral index of $\alpha=-0.8$ (synchrotron spectrum). Therefore, the 5\,GHz data are deeper than the data at the other four frequencies, which biases our sample to SFGs with a relatively steeper radio spectrum and causes the median radio spectral slope at observer-frame frequencies of 610--5,000\,MHz and 150--5,000\,MHz to be steeper than that of the synchrotron spectrum as shown in Section $\S$\ref{s:610_5000}. Because of this bias, we overestimate the fraction of SFGs with a low-frequency radio spectrum significantly flatter than that at high frequency shown in Figure~\ref{f:en1_gmrt_lofar_vla_SED_nml_restframe}.

Despite these biases, we find that the radio spectrum is flatter if we include a lower frequency dataset when measuring the two-point radio spectral indices between observer-frame frequencies of 150--610--1,250\,MHz, and 150--610--5,000\,MHz as shown in Sections $\S$\ref{s:uGMRT_1250MHz} and $\S$\ref{s:610_5000}. We will discuss the possible physical mechanisms that cause the radio spectrum of SFGs to be flatter at low frequency than at high frequency in Section $\S$\ref{s:flatter at low frequency}.

\subsection{Effects on $k$-correction}\label{s:k-correction}
For studies based on rest-frame radio luminosity (typically at $\sim$1.4\,GHz), such as FIRRC or radio-SFR relation, an appropriate $k$-correction is required. In this work, we take rest-frame luminosity at 1,250\,MHz ($L_{\rm 1.2\,GHz}$) to discuss how different assumptions in $k$-correction affect the estimation of $L_{\rm 1.2\,GHz}$. 

For SFGs at $0<z<1$, the appropriate radio spectral index used in $k$-correction is that at observed $\sim$610--1,250\,MHz, because the emission at rest-frame 1,250\,MHz is shifted to that frequency range. We compare our measured radio spectral indices at 610--1,250\,MHz with synchrotron spectrum ($\alpha=-0.8$), which is widely used in $k$-correcting the observed radio flux densities at $\sim$1.4 GHz in the literature \citep[e.g.,][]{Magnelli15, Hindson18, Delvecchio21}. We find that for the SFGs detected by GMRT-610MHz, 45\% of their 610--1,250MHz radio spectrum are significantly flatter than the synchrotron spectrum, i.e., $\alpha_{610}^{1250}-\sigma_{610}^{1250}>-0.8$. Therefore, 45\% of their $L_{\rm 1.2\,GHz}$ will be significantly overestimated if we use the synchrotron spectrum in $k$-correction. We point out that the fraction of 45\% is overestimated because of a relatively shallower uGMRT 1,250\,MHz data, which biases our sample to SFGs with a relatively flatter radio spectrum at 610--1,250\,MHz as we discussed in Section $\S$\ref{s:spectral indices}. 
However, the fraction of SFGs in which the low-frequency radio spectrum is flatter than the synchrotron spectrum increases with decreasing frequency. For instance, 93\% of SFGs in our sample have a radio spectrum at observed 150--1,250\,MHz that is significantly flatter than the synchrotron spectrum. Therefore, the individually measured radio spectral index is essential to $k$-correction in studies based on the rest-frame radio luminosity, especially for SFGs at high-redshift.

\subsection{Radio spectral properties at observer-frame 150--5,000\,MHz}\label{s:physical mechanisms}
For radio-selected SFGs, we find that, on average, their radio spectrum slightly steepens with increasing stellar mass at any frequency ranges analysed in this work. However, we only find that at $\nu_{\rm obs}\la$1\,GHz, the radio spectrum of SFGs slightly flattens with increasing optical depth. This results a positive correlation between spectral curvature ($\nu_{\rm obs}\sim$1\,GHz) and optical depth at $V$-band. At $\nu_{\rm obs}\la$1\,GHz, the spectral curvature ($\alpha_{150}^{400}-\alpha_{400}^{610}$) is not significantly correlated with physical properties of SFGs, which has been reported in \cite{Rivera17} but at observer-frequency of 150--325--1,400\,MHz. On average, the radio spectrum we measured at $\nu_{\rm rest}\la$1\,GH is flatter than that at the high frequency reported in the literature \citep[e.g.,][]{Delhaize17, An21} and the synchrotron spectrum. Here we discuss the possible explanations of these correlations.

\subsubsection{Radio spectrum steepens with increasing stellar mass}\label{s:alpha_stellar mass}
At the observer-frame frequencies of 150--5,000\,MHz, we find that the radio spectrum slightly steepens with increasing stellar mass.
Although this trend is statistically weak with a Spearman's rank correlation coefficient of $\rho_{\rm s}\sim-0.12$, it has been reported in our previous work that studied radio spectral properties at the observer-frame frequencies of 1.3--3\,GHz by combining the MeerKAT 1.3\,GHz and VLA 3\,GHz data in the COSMOS field \citep{An21} and in \cite{Heesen22} that studied nearby galaxies at $\sim$144-1,400\,MHz. In \cite{An21}, we noted that this correlation could be explained by age-related synchrotron loss.

For the relativistic CR electron, the critical frequency, at which the electron emits most of its energy, is proportional to the square of its energy and the strength of the perpendicular component of the magnetic field, i.e, 
\begin{eqnarray} \label{e:equation1}
{\nu_{\rm c}}\propto E^{2} B_{\bot}.
\end{eqnarray}
If we only take synchrotron radiation into account, the lifetime of a CR electron is 
\begin{eqnarray} \label{e:equation2}
{t_{\rm sy}}\propto {\nu_{\rm c}}^{-0.5} B_{\bot}^{-1.5}.
\end{eqnarray}
Therefore, high-energy CR electrons lose their energy faster than low-energy CR electrons. Similarly, the inverse Compton loss of these CR electrons is also greater at high frequency \citep{Pacholczyk70}. For SFGs, fresh CR electrons are constantly injected into the star-forming regions, which keeps the radio spectral slope constant. We therefore proposed in \cite{An21} that the slightly increased ratio of aged and young relativistic CR electrons with increasing stellar mass could be a physical cause leading to radio spectrum steepening for massive SFGs. Accordingly, the steepening of radio spectrum with increasing stellar mass is also expected to be intrinsically weak.

\subsubsection{Radio spectrum flattens with increasing optical depth at $\nu\la$1\,GHz}\label{s:optical depth}

At $\nu_{\rm obs}\la$1\,GHz, the radio spectrum of SFGs slightly flattens with increasing optical depth at $V$-band. However, this correlation becomes weak and even negligible when we include the high-frequency uGMRT 1,250\,MHz or VLA 5\,GHz data, although the sample size of SFGs with 1,250\,MHz or 5\,GHz detection is very limited because of the shallow or small coverage of the data. 

Although this trend is also statistically weak with a $\rho_{\rm s}\sim0.25$, similar trends have been reported in the literature. \cite{Condon91} and \cite{Murphy13} found that, for local ULIRGs, the radio spectra flatten with increasing compactness and silicate optical depth. These correlations might be a result of increased free–free absorption arising from more deeply embedded star formation as suggested by \cite{Murphy13}. Although the dust in these star-formation regions traces the cold gas that is mostly dominated by cool neutral and atomic hydrogen, the density of ionizing photons (including the free electrons, $n_{\rm e}$) of the ISM surrounding the dust is also high because of a high SFR. Therefore, the observed radio continuum spectrum is affected more by free-free absorption for the SFGs that have a relatively higher optical depth at $V$-band because: 
\begin{eqnarray} \label{e:equation3}
\kappa \propto n_{e}^2\,T_{e}^{-1.35}\,\nu^{-2.1}, 
\end{eqnarray}
where $\kappa$ is the free-free absorption coefficient and $T_{\rm e}$ is the electron temperature of the \HII\ emitting region \citep{Condon92}. However, for the integrated radio spectrum of SFGs, its association with the star-forming regions is smoothed, which explains the statistically weak correlation.

The Equation~\ref{e:equation3} and other theoretical models, as well as previous observations, show that the effect of thermal free-free absorption increases at low frequency \citep{Condon92,McDonald02, Tingay04, Murphy09, Clemens10, Lacki13, Kapinska17, Galvin18, Klein18, Dey22}, which explains why we only find the correlation between optical depth at $V$-band and radio spectrum at $\nu_{\rm obs}\la$1\,GHz.

\subsubsection{Radio spectrum is flatter at low frequency than at high frequency}\label{s:flatter at low frequency}

As discussed in Section $\S$\ref{s:spectral indices}, our measurements and the comparison between our measurements and radio spectral index at $\sim$150--5,000\,MHz in the literature show that, on average, the radio spectrum of SFGs is flatter at low frequency than at high frequency. However, when we investigate the measured radio spectral index as a function of rest-frame frequency (redshift) at either frequency range, i.e., 150-610\,MHz (Figure~\ref{f:en1_gmrt_lofar_deep_matched_alpha_z_Ms_final_conc}), 400--610\,MHz (Figure~\ref{f:en1_gmrt_deep_lofar_ugmrt_matched_alpha_z_Ms_final_curve_color_1}), or 610--1,250\,MHz (Figure~\ref{f:en1_gmrt_deep_1p2}), we only find a very weak or even negligible correlation because of the large scatter of measured radio spectral indices at any given rest-frame frequencies. Therefore, although the measured spectral indices sample different parts of the spectrum because of different redshift of SFGs (Figure~\ref{f:en1_gmrt_lofar_deep_matched_alpha_z_Ms_final_conc}, Figure~\ref{f:en1_gmrt_deep_lofar_ugmrt_matched_alpha_z_Ms_final_curve_color_1}, and Figure~\ref{f:en1_gmrt_deep_1p2}), the correlations between radio spectral index and stellar mass, and optical depth at $V$-band are less affected.

We also find a positive correlation between stellar mass and optical depth at $V$-band with a $\rho_{\rm s}\sim0.16$, which is expected because massive galaxies tend to be more dusty \citep{daCunha10, Kennicutt12, An14, Calura17, An17}. If this is a primary relation of SFGs, their radio spectral index's correlations with stellar mass and optical depth at $V$-band would be weakened. This further suggests that the underlying physical mechanisms that derive the two correlations are different as discussed in  Section $\S$\ref{s:alpha_stellar mass} and  Section $\S$\ref{s:optical depth}.

The correlations of SFGs' radio spectral slope with the stellar mass and with the optical depth at $V$-band suggest that both age-related synchrotron loss of CR electrons and thermal free-free absorption could be the physical causes of why the radio spectrum is flatter at low frequency than at high frequency. Other mechanisms, such as thermal emission, synchrotron self-absorption, bremsstrahlung and ionisation losses, and the Razin-Tsytovich effect, may also flatten the radio spectrum at low frequency. However, the thermal fraction is expected to be $\la$10\% at $\nu_{\rm obs}\la$1\,GHz \citep{Condon92, Klein18}. Therefore, thermal emission is an insufficient explanation for the galaxies with radio spectra deviating significantly from the injection spectrum (Figure~\ref{f:en1_gmrt_lofar_deep_matched_alpha_z_Ms_final_conc}).
Synchrotron self-absorption requires extremely high brightness temperatures in order to be significant, and so its impact in SFGs is expected to be weak \citep[e.g.,][]{Tingay04,Clemens10}. Bremsstrahlung and ionisation losses may dominate over synchrotron losses at $\nu_{\rm obs}\la$1\,GHz in less-luminous radio galaxies \citep{Hindson18}, these effects only become important for dense starburst regions that have a $\sim$1\,mG magnetic field, which is stronger than that of typical SFGs \citep{Murphy09}. In addition, for galaxies detected at LOFAR 150\,MHz, the SFR to radio luminosity relation also appears to hold, which suggests that these two mechanisms may have played a minor role in flattening the radio spectrum of SFGs at low frequency \citep{Smith21, Heesen22}.  
The Razin-Tsytovich effect can suppress synchrotron emission below the Razin frequency,
\begin{eqnarray} \label{e:equation4}
\nu_{\rm R} \propto n_{e}B^{-1}.
\end{eqnarray}
This effect may dominate over internal free-free absorption when the thermal gas is considerably hotter than 10$^4$\,K, which is rarely attained by normal SFGs \citep[e.g.,][]{Condon91,Lacki13, Chyzy18}. These leave the thermal free-free absorption and CR electrons energy loss are the most important causes of why the radio spectrum is flatter at low frequency than at high frequency. The former flattens the radio spectrum at low frequency, while the latter steepens the radio spectrum at high frequency \citep[e.g.,][]{Condon91, McDonald02, Murphy09, Clemens10, Lacki13, Murphy13, Marvil15, Kapinska17, Chyzy18, Galvin18, Klein18, Dey22, Heesen22}.

The two-point radio spectral index we measure represents the averaged radio spectral slope between the two frequencies. The discussions in Section $\S$\ref{s:alpha_stellar mass} and Section $\S$\ref{s:optical depth} suggest that both the flattening of the radio spectrum at low frequency and the steepening of the radio spectrum at high frequency are gradual. In addition, as shown by Equation \ref{e:equation1} to \ref{e:equation4}, the magnetic field ($B$), column density of free electrons ($n_{e}$), and electron temperature ($T_{\rm e}$) in the \HII\ region are the key features of the physical mechanisms that are responsible for the different radio spectral slopes at low and high frequency discussed above. Studies of radio spectra of nearby galaxies and theoretical simulations of CR electron propagation discussed the effects of these individual physical parameters as well as some other physical conditions, such as galactic wind \citep[e.g.,][]{Heesen16, Mulcahy18, Winner19, Werhahn21, Stein23}. 

In contrast, one advantage of our statistical studies is the ability to suggest possible causes of why the radio spectrum is flatter at low frequency than at high frequency. In particular, we find that the correlations between the SFGs' integrated radio spectrum and their astrophysical properties could lend support to CR electrons energy loss and thermal free-free absorption as two possible main mechanisms for our samples.

\section{Conclusion}\label{s:conclusion}
By combining high-sensitivity LOFAR 150\,MHz, uGMRT 400\,MHz and 1,250\,MHz, GMRT 610\,MHz and VLA 5\,GHz data in the ELAIS-N1 field, we measure the radio spectral indices between observer-frame frequencies of 150--5,000\,MHz and study their correlations with physical properties of SFGs. The main conclusions from our work are as follows.

1) We obtain $\sim$3,500 SFGs that have detections at both GMRT 610\,MHz and LOFAR 150\,MHz by removing AGN from the two radio samples. The median radio spectral index for these SFGs is $\alpha^{610}_{150}=-0.51\pm0.01$ with a scatter of $\sigma=0.3$. The range of the photometric redshift of these SFGs is $z=0.01-6.21$.

2) Because of the relatively low-sensitivity of uGMRT 400\,MHz data, we limit our sample to the SFGs with $S_{\rm 610\,MHz}>300\,\mu$Jy, and obtain the median spectral indies of $\alpha^{400}_{150}=-0.40\pm{0.02}$, $\alpha^{610}_{400}=-0.49^{+0.05}_{-0.04}$, and $\alpha^{610}_{150}=-0.42^{+0.02}_{-0.01}$.

3) The uGMRT 1,250\,MHz data are also relatively shallower compared with the GMRT-deep 610\,MHz and LOFAR 150\,MHz data. We obtain the median two-point spectral indices of $\alpha_{610}^{1250}=-0.56^{+0.04}_{-0.02}$ and $\alpha_{150}^{1250}=-0.48\pm0.01$ for the 315 SFGs that are brighter than 60\,$\mu$Jy at 1,250\,MHz. Although the shallow uGMRT 1,250\,MHz data bias our sample to SFGs with a relatively flatter radio spectrum at 150--1,250\,MHz (also 610--1,250\,MHz), we find that the radio spectrum is flatter if we include a lower frequency dataset when measuring the two-point radio spectral indices at observed 150--1,250\,MHz.

4) Although the VLA 5\,GHz data have very limited coverage, and its high-sensitivity biases the sample to SFGs with a relatively steeper radio spectrum, our measured radio spectral indices at 150--5,000\,MHz and the comparison between our measurements and the results in the literature show that, on average, the radio spectrum of SFGs is flatter at low frequency than at high frequency.

5) We correlate the measured radio spectral indices with the physical properties of SFGs and find that, on average, the radio spectrum slightly steepens with increasing stellar mass at the observer-frame frequencies of 150--5,000\,MHz. We suggest that the spectral ageing due to the energy loss of relativistic CR electrons via synchrotron and inverse Compton radiation could explain this trend. 

6) At $\nu_{\rm obs}\la$1\,GHz, we find that, on average, the radio spectrum slightly flattens with increasing optical depth at $V$-band of radio-selected SFGs. However, this correlation becomes weak or even negligible when we include the high-frequency uGMRT 1,250\,MHz or VLA 5\,GHz data. We discuss that thermal free-free absorption is a possible underlying physical mechanism of this correlation. 

7) Our statistical study of the correlation between integrated radio spectrum and physical properties of SFGs suggests the spectral ageing of CR electrons and thermal free-free absorption are two main possible physical mechanisms that cause the radio spectrum to be flatter at low frequency than at high frequency.

\section*{Acknowledgments}

We acknowledge Akriti Sinha for providing the uGMRT 1,250 MHz catalogue. FXA and MV acknowledge financial support from the Inter-University Institute for Data Intensive Astronomy (IDIA), a partnership of the University of Cape Town, the University of Pretoria, the University of the Western Cape, and the South African Radio Astronomy Observatory. FXA is grateful for support from the National Science Foundation of China (12303016, 12233005, 12073078, and 12173088). MV acknowledges the support from the South African Department of Science and Innovation’s National Research Foundation under the ISARP RADIOSKY2020 and RADIOMAP Joint Research Schemes (DSI-NRF Grant Numbers 113121 and 150551) and the SRUG HIPPO Projects (DSI-NRF Grant Numbers 121291 and SRUG22031677). PNB and RK acknowledge support from the UK STFC via grant ST/V000594/1.

We acknowledge the use of the ilifu cloud computing facility –www.ilifu.ac.za, a partnership of the University of Cape Town (UCT), the University of the Western Cape, the University of Stellenbosch, Sol Plaatje University, the Cape Peninsula University of Technology, and the South African Radio Astronomy Observatory. The ilifu facility is supported by contributions from the IDIA, the Computational Biology division at UCT, and the Data Intensive Research Initiative of South Africa (DIRISA).

\section*{Data Availability}
The LOFAR 150\,MHz data used for the analysis of this paper are publicly available at https://lofar-surveys.org/deepfields.html. The uGMRT 400\,MHz and 1,250\,MHz data are available in \cite{Chakraborty19} and \cite{Sinha23} respectively. The GMRT 610\,MHz and VLA 5\,GHz data used in this research will be shared upon reasonable request to the corresponding author. 



\bibliographystyle{mnras}
\bibliography{LOFAR_GMRT_Fangxia} 

\appendix
\section{GMRT-wide 610\,MHz as selection frequency}\label{s:gmrt_wide_select}

\subsection{Sample of SFGs of GMRT-wide 610\,MHz-detected sources}\label{s:sample_wide_select}
We also use the relatively shallower but wide GMRT 610\,MHz data from \cite{Ishwara20} to study radio spectral properties between the observer-frame frequencies of 150--610\,MHz for radio-selected SFGs. Due to the similar coverages of GMRT-wide and LOFAR 150\,MHz data, we use the sample of SFGs classified in \cite{Best23} as the parent sample in the following analyses. For the 6,400 GMRT 610\,MHz-detected sources \citep{Ishwara20}, 4,542 are within the region that has both LOFAR and multi-wavelength observations (Figure~\ref{f:coverage}). Among them, 4,070 (90\%) have a LOFAR 150\,MHz counterparts within 10$\arcsec$ and 1,857 are classified as SFGs in \cite{Best23}. In addition, 1,258/6,400 overlap with GMRT-deep detected sources, and 394 are classified as AGNs in \cite{Ocran20}. The overlap between the two AGN samples is 342. By removing these additional 52 AGN from the 1,857 SFGs classified in \cite{Best23}, we obtain the final sample of 1,805 SFGs.

\subsection{Radio spectral properties at 150--610\,MHz based on GMRT-wide sample}

\begin{figure}
\includegraphics[width=0.48\textwidth]{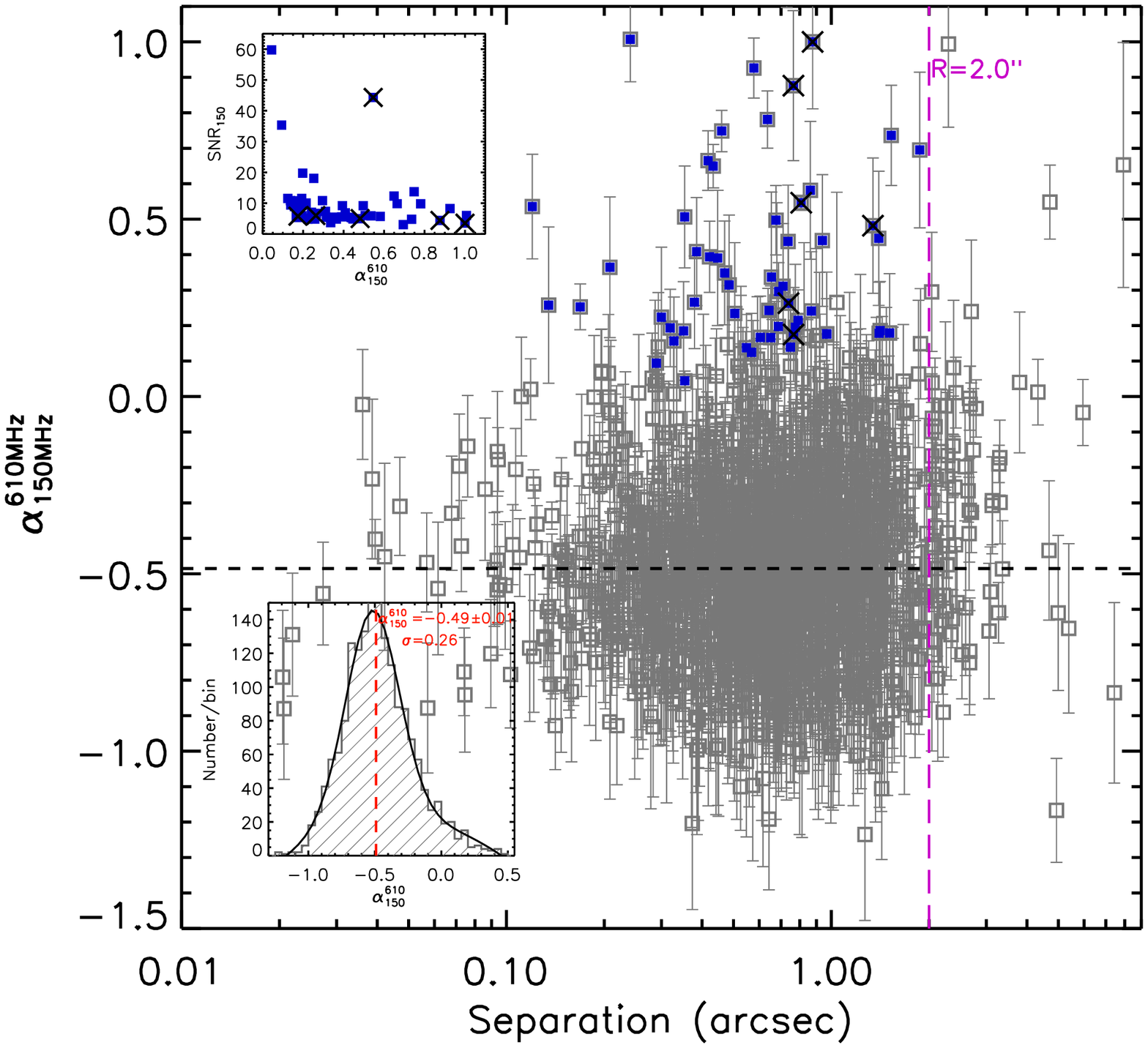}
\caption{Radio spectral index between the observer-frame frequencies of 150--610\,MHz plotted against the separation between LOFAR 150\,MHz and GMRT 610\,MHz detections for the 1,677 SFGs detected from GMRT-wide image data \citep{Ishwara20}. The purple dashed line marks the cross-match radius ($r_{150-610}=2\arcsec$) for identifying the LOFAR 150\,MHz counterparts of the GMRT 610\,MHz-detected sources. We mark the 43 sources with $r_{150-610}<2\arcsec$ and $\alpha_{150}^{610}-\sigma>0.0$ by blue squares. The top inset plot shows their SNR at 150\,MHz as a function of $\alpha_{150}^{610}$. Our visual inspection shows that six of these 43 sources are resolved as two components at LOFAR 150\,MHz, which are marked by black crosses. We remove these six sources from our sample and show the distribution of the radio spectral index for the remaining 1,597 SFGs with $r_{150-610}<2\arcsec$ in the bottom inset plot. The median radio spectral index between the observer-frame frequencies of 150--610\,MHz for these 610\,MHz-selected SFGs is $\alpha_{150}^{610}=-0.49\pm0.01$ with a scatter of $\sigma=0.26$.
}
\label{f:en1_gmrt_lofar_wide_matched_alpha}
\end{figure}

The measured radio spectral indices between the observer-frame frequencies of 150--610\,MHz of these 1,805 SFGs are shown in Figure~\ref{f:en1_gmrt_lofar_wide_matched_alpha}.
We also adopt $r_{150-610}=2\arcsec$ as the cut when cross-matching the GMRT-wide 610\,MHz and LOFAR 150\,MHz data as shown in Figure~\ref{f:en1_gmrt_lofar_wide_matched_alpha}. There are 48 sources with $r<2\arcsec$ that have $\alpha_{150}^{610}-\sigma_{150}^{610}>0.0$ ( highly flat radio spectrum between the observer-frame frequencies of 150-610\,MHz), and we show their SNR at 150\,MHz as a function of $\alpha_{150}^{610}$ in the top inset plot in Figure~\ref{f:en1_gmrt_lofar_wide_matched_alpha}. We visually inspect both LOFAR 150\,MHz and GMRT 610\,MHz images and find that six of these 48 sources are resolved as two components at LOFAR 150\,MHz. We, therefore, remove these six sources in the following analyses. Besides, as shown in the top inset plot of Figure~\ref{f:en1_gmrt_lofar_wide_matched_alpha}, the sources with an highly flatter radio spectrum at 150--610\,MHz have lower SNR at 150\,MHz. For the remaining 1,711 SFGs with $r_{150-610}<2\arcsec$, we obtain a median radio spectral index of $\alpha_{150}^{610}=-0.49\pm0.01$ with a scatter of $\sigma=0.26$, which are well-consistent with the results based on the GMRT-deep data (Section $\S$\ref{s:gmrt_select}). We note that 396 of these SFGs are also included in the GMRT-deep sample shown in Figure~\ref{f:en1_gmrt_lofar_deep_matched_alpha_z_Ms_final_conc}.

\begin{figure*}
\centering
\includegraphics[width=0.98\textwidth]{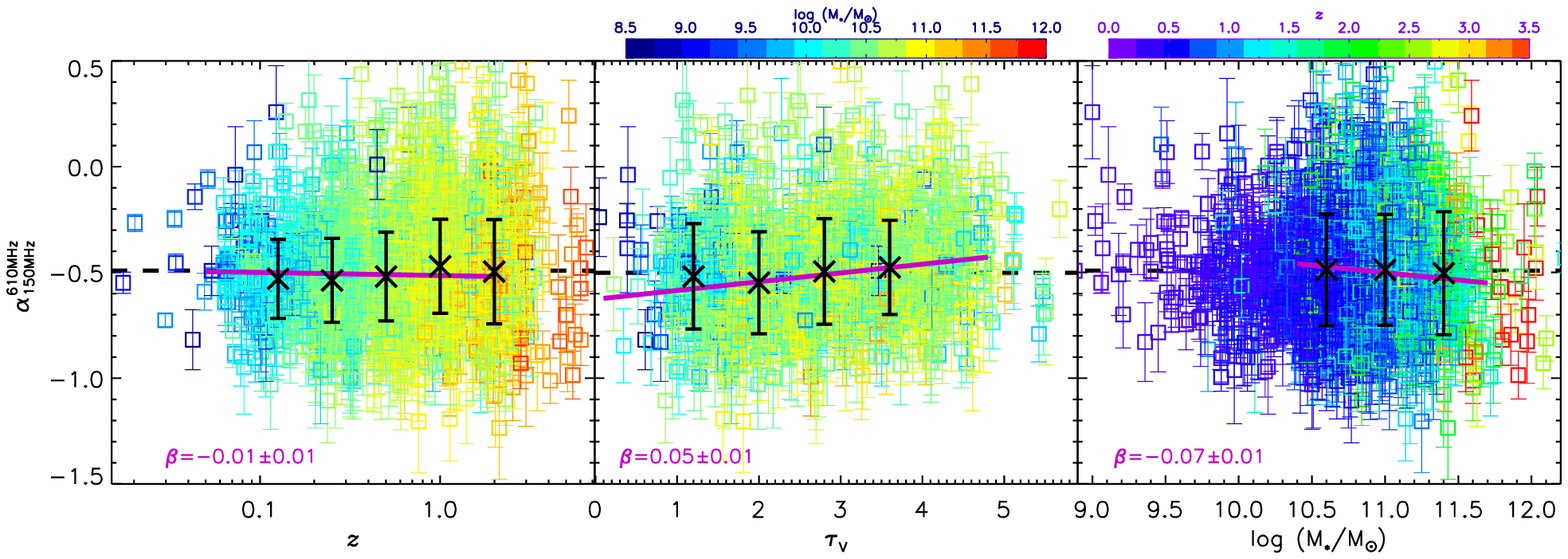}
\caption{Radio spectral index between observer-frame frequencies of 150--610\,MHz as functions of redshift ({\it left}), optical depth at $V$-band ({\it middle}), and stellar mass ({\it right}) for the SFGs detected from GMRT-wide \citep{Ishwara20}. The colour of the symbols represents the redshift in the right panel and the stellar mass in the other two panels. To reduce the effect from the incompleteness of our radio flux-limited sample, we limit the SFGs to $z<2$ and log $(M_{*}/\Msun)>10.4$ when fitting the correlations between $\alpha_{150}^{610}$ and redshift, and stellar mass respectively. On average, the radio spectral slope of SFGs at observed 150--610\,MHz flattens with increasing optical depth and steepens with increasing stellar mass, although we do not find a significant correlation between $\alpha_{150}^{610}$ and redshift (rest-frame frequency).}
\label{f:en1_gmrt_lofar_wide_matched_alpha_z_Ms_final_conc}
\end{figure*}

For the 1,711 GMRT-wide-detected SFGs, 1,296 with $z<1.5$ have estimated optical depth at $V$-band, $\tau_{V}$, from \cite{Smith21}. As shown in Figure~\ref{f:en1_gmrt_lofar_wide_matched_alpha_z_Ms_final_conc}, on average, the radio spectrum at observed 150--610\,MHz for the GMRT-wide-selected SFGs also slightly flattens with increasing optical depth and steepens with increasing stellar mass. However, we do not find the weak correlation between $\alpha_{150}^{610}$ and redshift (rest-frame frequency). This might be caused by the sample of SFGs being not as clean as the one based on GMRT-deep-detected sources (Section $\S$\ref{s:gmrt_select}). Overall, the radio spectral properties of GMRT-wide selected SFGs are consistent with those based on the GMRT-deep selected sample. 

%
%

%
%

\section{uGMRT 400\,MHz as the selection frequency}\label{s:ugmrt_select}
%
%
\begin{figure*}
\centering
\includegraphics[width=0.98\textwidth]{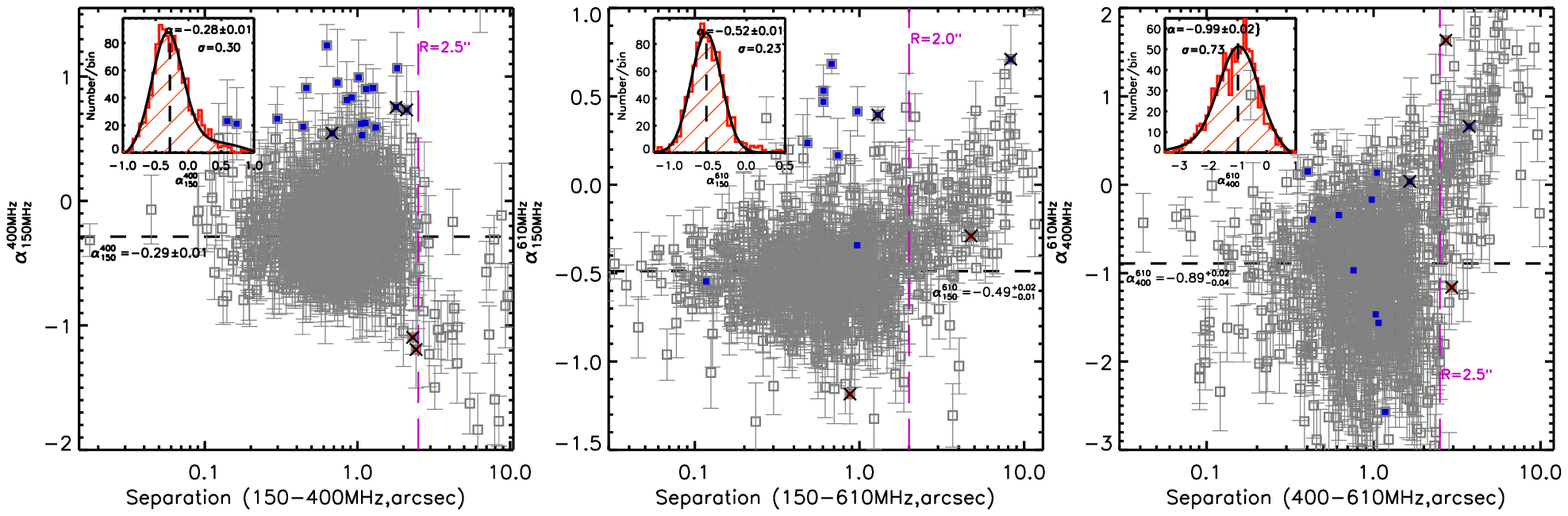}
\caption{Radio spectral index between the observer-frame frequencies of 150--400\,MHz ({\it left}), 150--610\,MHz ({\it middle}), and 400--610\,MHz ({\it right}) plotted against the separation between each pair of frequencies for the uGMRT 400\,MHz-detected radio sources. {\it Left:} The purple dashed line marks the cross-match radius of $r_{150-400}=2\farcs5$, which is chosen by the distribution of $\alpha_{150}^{400}$ for the 1,194 uGMRT 400\,MHz SFGs that have a LOFAR 150\,MHz counterparts within 10$\arcsec$. The blue squares are the 20 SFGs with $r_{150-400}<2\farcs5$ and $\alpha_{150}^{400}-\sigma>0.3$, while the red squares are the two SFG with $\alpha_{150}^{400}+\sigma<-0.8$. Our visual inspection shows that five of them are resolved at LOFAR 150MHz, which are marked by black crosses. We remove these five sources and show the distribution of $\alpha_{150}^{400}$ for the remaining 1,158 SFGs with $r_{150-400}<2\farcs5$ in the left-top inset plot. The black dashed line marks their median spectral index of $\alpha_{150}^{400}=-0.28\pm0.01$ with a scatter of $\sigma=0.30$. {\it Middle:} For these 1,194 uGMRT 400\,MHz SFGs, 1,038 also have a GMRT 610\,MHz counterpart within 10$\arcsec$. After removing the sources that are resolved at LOFAR 150\,MHz, we obtain a sample of 896 SFGs with $r_{150-610}<2\arcsec$ and a median spectral index of $\alpha_{150}^{610}=-0.52\pm0.01$ with a scatter of 0.23 as shown in the left-top inset plot. {\it Right:} We also apply a cross-match radius of $r_{400-610}=2\farcs5$ (purple dashed line) and obtain a sample of 926 SFGs that have detections at both uGMRT 400\,MHz and GMRT 610\,MHz. The median spectral index of these SFGs is $\alpha_{400}^{610}=-0.99\pm0.02$ with a scatter of $\sigma=0.73$ as shown in the left-top inset plot.} 
\label{f:en1_uGMRT_GMRTdeep_LOFAR}
\end{figure*}

\subsection{Sample of SFGs of uGMRT 400\,MHz-detected sources}\label{s:sample_ugmrt_select}
We use the uGMRT 400\,MHz-detected sources from \cite{Chakraborty19} to study the radio spectrum between observer-frame frequencies of 150--400\,MHz and 400--610\,MHz respectively. Because of the large coverage and high sensitivity of LOFAR 150\,MHz data, we also use the sample of SFGs classified in \cite{Best23} as the parent sample in this section. For the 2,528 uGMRT 400\,MHz radio sources from \cite{Chakraborty19}, 2,219 (88\%) have a LOFAR 150\,MHz counterpart within 10$\arcsec$. Among them, 1,255 are classified as SFGs in \cite{Best23}. As shown in Figure~\ref{f:coverage}, the coverages of uGMRT 400\,MHz and GMRT-deep 610\,MHz are slightly offset. Within the overlapped region, 88\% (2,110/2,386) of uGMRT 400\,MHz sources have a 610\,MHz counterpart within 10$\arcsec$. Among them, 1,180 are within 2$\farcs$5 and 401/1,180 are classified as AGNs in \cite{Ocran20}. The overlap between the two AGN samples from \cite{Best23} and \cite{Ocran20} is 340. Therefore, we remove the additional 61 AGN from the 1,255 SFGs classified in \cite{Best23} and obtain the final sample of 1,194 SFGs.

\subsection{Radio spectral indices at observed 150--400\,MHz and 400--610\,MHz}
We show the radio spectral index between observer-frame frequencies of 150--400\,MHz as a function of the separation between LOFAR 150\,MHz and uGMRT 400\,MHz detections for these 1,194 SFGs in the left panel of Figure~\ref{f:en1_uGMRT_GMRTdeep_LOFAR}. According to the distribution of $\alpha_{150}^{400}$, we choose $r_{\rm 150-400}=2\farcs5$ as the cut when cross-matching the uGMRT 400\,MHz and LOFAR 150\,MHz sources. There are 20 SFGs with $r_{\rm 150-400}<2\farcs5$ that have $\alpha_{150}^{400}-\sigma_{150}^{400}>0.3$, which means that they have highly flat radio spectra at observed 150--400\,MHz. Our visual inspection shows that three of them are resolved as two components at LOFAR 150\,MHz. There are also two SFGs with $\alpha_{150}^{400}+\sigma_{150}^{400}<-0.8$, i.e., with highly steep radio spectra at observed 150--400\,MHz. We find that both of them are resolved at LOFAR 150\,MHz. We, therefore, remove these five sources from our sample and show the distribution of $\alpha_{150}^{400}$ for the remaining 1,158 SFGs with $r_{\rm 150-400}<2\farcs5$ in Figure~\ref{f:en1_uGMRT_GMRTdeep_LOFAR}. The median radio spectral index between observer-frame frequencies of 150--400\,MHz is $\alpha_{150}^{400}=-0.28\pm0.01$ with a scatter of $\sigma=0.30$.

For the 1,194 SFGs that have both uGMRT 400\,MHz and LOFAR 150\,MHz detections, 1,038 also have a GMRT 610\,MHz counterpart within 10$\arcsec$. We show their radio spectral indices between observer-frame frequencies of 150--610\,MHz as a function of the separation between LOFAR 150\,MHz and GMRT 610\,MHz detections in the middle panel of Figure~\ref{f:en1_uGMRT_GMRTdeep_LOFAR}. Among them, 896 have $r_{\rm 150-610\,MHz}<2\arcsec$ after we remove the sources that are resolved at LOFAR 150\,MHz. The median $\alpha_{150}^{610}$ for these 896 uGMRT 400\,MHz-selected SFGs is $\alpha_{150}^{610}=-0.52\pm0.01$ with a scatter of $\sigma=0.23$, which is consistent with the results based on GMRT 610\,MHz-detected SFGs (Section $\S$\ref{s:gmrt_select} and Appendix~\ref{s:gmrt_wide_select}).

We also show the radio spectral index between observer-frame frequencies of 400--610\,MHz as a function of the separation between uGMRT 400\,MHz and GMRT 610\,MHz detections for the 1,038 uGMRT 400\,MHz-selected SFGs that have a GMRT 610\,MHz counterpart within 10$\arcsec$ in the right panel of Figure~\ref{f:en1_uGMRT_GMRTdeep_LOFAR}. According to the distribution of $\alpha_{400}^{610}$, we choose $r_{\rm 400-610\,MHz}=2\farcs5$ as the cut when cross-matching uGMRT 400\,MHz and GMRT 610\,MHz sources. There are 926 uGMRT 400\,MHz-selected SFGs that have $r_{\rm 400-610\,MHz}<2\farcs5$ after we remove the SFGs that are resolved at LOFAR 150\,MHz. The median spectral index is $\alpha_{400}^{610}=-0.99\pm0.02$ with a scatter of $\sigma=0.73$. 
The differences in the spectral slope above and below 400\,MHz occur because of the relatively shallower uGMRT 400\,MHz data, i.e., the radio spectral indices we measured are flatter over the 150--400\,MHz and steeper over the 400--610\,MHz (Eddington bias). 

\section{Cross-matches between uGMRT 1,250\,MHz and other radio data}\label{s:1250_select}
%
%
\begin{figure*}
\centering
\includegraphics[width=0.98\textwidth]{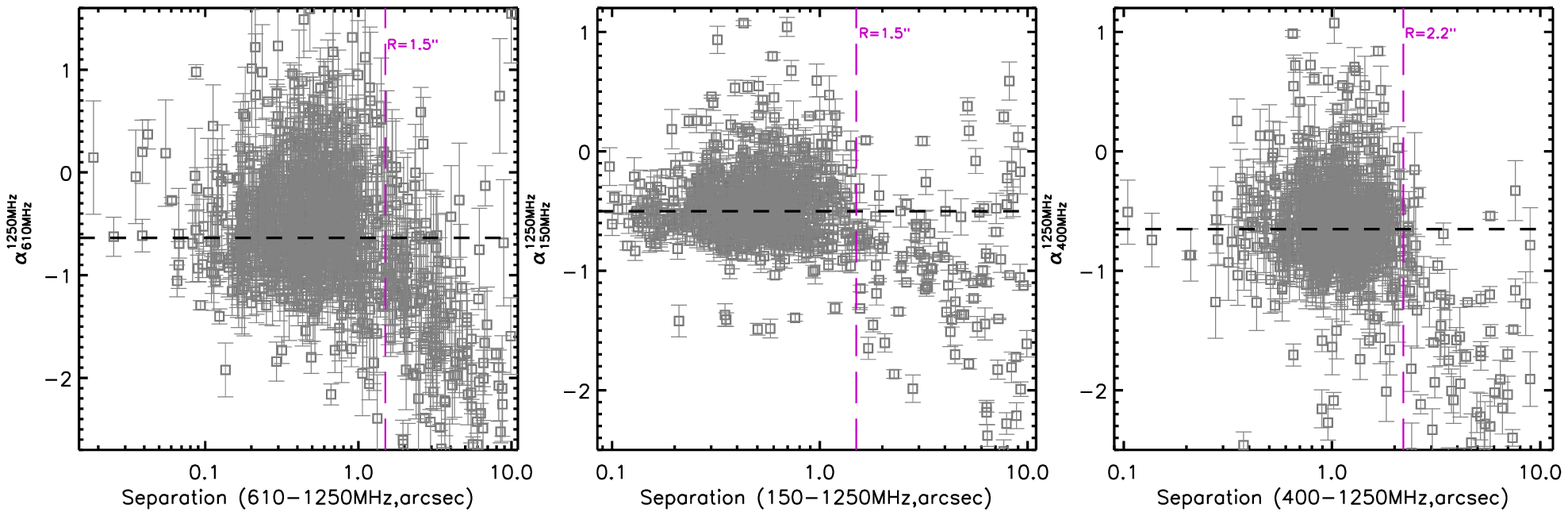}
\caption{Radio spectral index at observed 610--1,250\,MHz ({\it left}), 150--1,250\,MHz ({\it middle}), and 400--1,250\,MHz ({\it right}) plotted against the separation between each pair of frequencies for the uGMRT 1,250\,MHz-detected radio sources. The purple dashed lines represent the cross-match radius between the two corresponded datasets, which are chosen by the distributions of two-point radio spectral index.}
\label{f:en1_uGMRT_1p2_GMRTdeep_LOFAR_matched}
\end{figure*}

To study radio spectral properties at 150--1,250\,MHz, we first choose the cross-matching radii between uGMRT 1,250\,MHz-detected sources \citep{Sinha23} and other radio data used in this work. For the 1,086 uGMRT 1,250\,MHz-detected sources from \cite{Sinha23}, 94\%, 87\%, and 87\% of them have the GMRT-deep 610\,MHz, LOFAR 150\,MHz, and uGMRT 400\,MHz counterparts within 10$\arcsec$ respectively. We plot the measured two-point radio spectral indices against the separation between each pair of frequencies for the uGMRT 1,250\,MHz-detected sources in Figure~\ref{f:en1_uGMRT_1p2_GMRTdeep_LOFAR_matched}. According to the distributions of two-point radio spectral indices, we choose $r=1\farcs5$ as the cross-match radii between uGMRT 1,250\,MHz and GMRT 610\,MHz, and LOFAR 150\,MHz. The cross-match radius between uGMRT 1,250\,MHz and 400\,MHz data is slightly larger ($r=2\farcs$2) because of a relatively large astrometric offset between the two datasets as shown in Figure~\ref{f:astrometry_lofar_gmrt}.


\bsp	
\label{lastpage}
\end{document}